\documentclass[prb, twocolumn]{revtex4}
\usepackage{graphicx}
\newcounter{subfigure}
\begin{document}
\title{\flushleft{Dynamics of propagating turbulent pipe flow
    structures. Part I: Effect of drag reduction by spanwise wall oscillation }}
% \newline \newline \hspace{2em} \normalsize  \textrm{Andrew Duggleby and Kenneth S. Ball} \newline \hspace{2em}
%\textrm{Department of Mechanical Engineering, Virginia Polytechnic Institute and State University, Blacksburg, Virginia  24061} \newline  \textrm{Received}}
\date{\today}
\author{A. Duggleby}
\author{K. S. Ball}
\author{M. R. Paul}
\affiliation{Department of Mechanical Engineering, Virginia Polytechnic Institute and State University Blacksburg, Virginia  24061}
\email{duggleby@vt.edu}

\begin{abstract}
\noindent The results of a comparative analysis based upon a Karhunen-Lo\`{e}ve
 expansion of turbulent pipe flow and
drag reduced turbulent 
pipe flow by spanwise wall oscillation  are presented. 
The turbulent flow is generated by a direct numerical simulation at a
Reynolds number $Re_\tau = 150$.  The spanwise wall oscillation is
imposed as a velocity boundary condition with an amplitude of $A^+ =
20$ and a period of $T^+ = 50$.  The wall oscillation results in a
27\% mean velocity increase when the flow is driven by a constant
pressure gradient.  The
peaks of the Reynolds stress and root-mean-squared velocities  shift
away from the wall and the Karhunen-Lo\`{e}ve dimension of the
 turbulent attractor is reduced from $2453$ to
$102$.  The coherent vorticity
structures are pushed away from 
the wall into higher speed flow, causing an increase of their advection
speed of 34\% as determined by a normal
speed locus.  This increase in advection speed gives the
propagating waves less time to interact with the roll modes.  This 
leads to less energy transfer and a shorter lifespan of the propagating structures, and thus less
Reynolds stress production which results in drag reduction.

\end{abstract}

\maketitle

\section{INTRODUCTION}
In the last decade a significant amount of work has been performed
investigating the structure of  wall bounded turbulence, with aims of
understanding its self-sustaining  nature and discovering methods of
control. \cite{panton_overview}  One of the greatest potential benefits for controlling
turbulence is drag reduction.  As the mechanics of the different
types of drag reduction are studied, most explanations of the mechanism revolve around controlling the streamwise
vortices and low speed streaks. \cite{karniadakis_choi, choi_nature06}
 One such method of achieving drag reduction is spanwise wall  
oscillation, first discovered by Jung et al. \cite {jung} in 1992,
and later confirmed both  numerically
\cite{quadrio, choi, jung, zhou_dongmei, quadrio2} and experimentally,
\cite{choiGraham, trujillo, choiEXP,  skandaji,laadhari}
 to reduce drag on the order of
45\%.  The prevalent theory of the mechanism behind this was
developed using
direct numerical simulation of a turbulent channel flow by Choi et al. \cite{choi} and
experimentally confirmed by Choi and Clayton, \cite{choi_clayton} showing
that the spatial correlation between the streamwise vortices and the
low speed streaks are modified so that high speed fluid is ejected
from the wall, and low speed fluid is swept towards the wall.
Even though this proposed mechanism describes the near-wall dynamics
that govern the drag reduction, questions behind the global dynamics have
not been sufficiently resolved.  What is the effect (if any) on the
outer region of the flow?  How do the coherent structures of the flow
near the wall adjust with the spanwise oscillations?  What is the
effect on the interactions between the inner and outer layers?

One manner in which to address these questions is through
direct numerical simulation (DNS) of turbulence.  As supercomputing resources increase, DNS continues to provide an information rich testbed to
investigate the dynamics and  mechanisms behind
turbulence and turbulent drag reduction.  DNS resolves all the
scales of turbulence without the need of a turbulent model and
provides a three dimensional time history of the entire flow
field.  One of the methods used for mining the information generated
by DNS
is the Karhunen-Lo\`{e}ve (KL) decomposition, which extracts coherent structures
from the eigenfunctions of the two-point spatial correlation
tensor. This allows a non-conditionally based investigation that
takes advantage of the richness of DNS.  The utility of this method is evident in the knowledge it has produced so far,
such as the discovery of propagating structures (traveling waves) of constant phase velocity that
trigger bursting and sweeping events. \cite{sirovich1, sirovich2,
  ball}   These studies in turn have lead to a new class of methods
for achieving drag reduction through wall imposed traveling waves.
\cite{itoh, karniadakis_choi}  Another study using KL decomposition examined the
energy transfer path from the applied pressure gradient to the flow 
through triad interaction of structures, \cite{webber1, webber2}
explaining the dynamical interaction between the KL modes.  In the realm
of control, KL methods have been used to produce drag reduction in a
turbulent channel by phase randomization of the structures \cite{handler} and
to understand the effect of drag reduction by controlled wall normal
suction and blowing.\cite{prabhu}  In the present study, the KL
framework is used to examine the differences in the turbulent
structures and dynamics 
between 
turbulent pipe flow with and without spanwise wall oscillation.

For this comparative analysis, turbulent pipe flow was chosen as opposed to turbulent
channel flow because of its industrial relevance and experimental accessibility.  The
main difference between pipe and channel flows is that in turbulent pipe flow the mean flow profile
exhibits a logarithmic profile that overshoots the theoretical profile
at low Reynolds numbers, whereas in turbulent channel flow it does not.
\cite{patel_head, durst}  Secondly, pipe flow differs
from channel flow because pipe flow is linearly stable to infinitesimal
disturbances where channel flow is linearly
unstable above a critical Reynolds number. \cite{orszag_patera,
  oSullivan_breuer}  A significant computational difference between pipe and
channel flow is the presence of a numerical
difficulty introduced by the singularity in polar-cylindrical
coordinates at the pipe centerline, which has limited the number of DNS studies in turbulent pipe
flow.  \cite{eggels, verzicco, fukagata, ma, loulou, fatica1997}
 Similarly, many KL studies have been performed in a turbulent channel
 flow, but to the best
of our knowledge the work presented here is the first to extend the KL method to spanwise wall oscillated turbulent
pipe flow.

In previous
work a DNS
of turbulent pipe flow for $Re_\tau=150$ was benchmarked and its
KL expansion was reported, forming the baseline for this study.\cite{duggleby_JOT}  
Similar structures to those
of turbulent channel flow \cite{sirovich1, sirovich2, ball} were found, including the presence of propagating modes. 
These propagating modes are characterized by a nearly constant phase speed and are
responsible for the Reynolds stress production as they interact and
draw energy from the roll
modes (streamwise vortices). \cite{sirovich1}  Without this
interaction and subsequent energy transfer, the propagating waves
decay quickly, reducing the total Reynolds stress of the flow.\cite{kerswell}   As shown by
Sirovich et al., \cite{sirovich1, sirovich2} the interaction between
the propagating waves and the roll modes  occurs by the propagating waves
forming a coherent oblique plane wave packet.  This wave packet interacts with the roll
modes, and when given enough
interaction time, the roll mode is destabilized eventually resulting in a bursting
event. \cite{sirovich1} It is in this bursting event that the energy
is transferred from the rolls to the propagating waves.\cite{webber1}
 In this paper we show that in the presence of spanwise wall
oscillation these propagating modes are pushed away from the wall into
higher speed flow.  This causes the propagating modes to advect faster, giving them
 less time to interact with the roll modes.  This
leads to reduced energy transfer that occurs less often, and yields lower
Reynolds stress production, which ultimately results in drag reduction.

\section{NUMERICAL METHODS}

We use a globally high order spectral element Navier-Stokes algorithm 
to generate turbulent data for pipe flow driven by
a mean streamwise pressure gradient. \cite{fischer_patera, tufo}  The non-dimensional equations
governing the fluid are

\begin{eqnarray}
\label{Navier_Stokes}
& \partial_t \mathbf{U}+\mathbf{U}\cdot \nabla
\mathbf{U}=-\nabla P+\mathrm{Re}_\tau^{-1} \nabla^2 \mathbf {U} \\
\label {continuity}
& \nabla \cdot \mathbf{U} = 0,
\end{eqnarray}
where $\mathbf{U}$ is the velocity vector, $\mathrm{Re}_\tau$ is the
Reynolds number, and $P$ is the pressure.  The velocity is non-dimensionalized by the wall shear
velocity $U_\tau=\sqrt{\tau_w / \rho}$ where $\tau_w$ is the wall
shear stress and $\rho$ is the density.  The Reynolds number is $\mathrm{Re}_\tau= U_\tau R / \nu  =
150$, where $R$ is the
radius of the pipe, and $\nu$ is the kinematic viscosity.  When
non-dimensionalized with the centerline velocity, the Reynolds number
is $Re_c \approx 4300$.   Two simulations
were performed, one with and one without spanwise wall oscillation.
In a pipe, the spanwise direction corresponds to the azimuthal
direction, so the oscillation is about the axis of the pipe.
Each case was run for $t^+= U_\tau^2 t / \nu=16800$ viscous time
units.  In the oscillated case, the simulation was
performed with an azimuthal velocity wall boundary condition $v_\theta (r=R, \theta, z)=A^+ \sin ( 2 \pi t /T^+)$ of
amplitude $A^+= A / U_\tau =20$ and period $T^+= U_\tau^2 T / \nu =50$
 with $(r, \theta,z)$
being the radial, azimuthal, and streamwise coordinates respectively.  This is not intended to
be a parametric study and the
amplitude and period were chosen to achieve maximum possible drag reduction
before relaminarization occurred.  

\begin{figure}
\centering
\includegraphics[width=2 in]{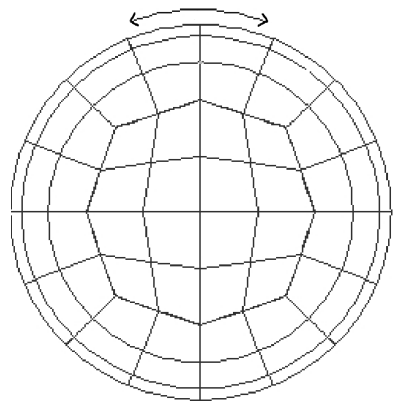}
\caption{A cross-section of a spectral element grid used for the pipe
  flow simulation.  The arrow denotes the direction of the spanwise
  (azimuthal) oscillation about the axis of the pipe.}
\label{grid}
\end{figure}

To solve equations \ref{Navier_Stokes} and \ref{continuity} we use a numerical algorithm employing a 
geometrically 
flexible yet exponentially convergent spectral element discretization in
space.  The spatial domain is subdivided into elements, each containing a high-order
(12th order) Legendre Lagrangian interpolant. \cite{lottes}  The
spectral element algorithm elegantly avoids the numerical singularity found in polar-cylindrical
coordinates at the origin, as seen in Figure \ref{grid}.  The
streamwise direction contains 40
spectral elements over a length of 10 diameters.  The effective
resolution of the flow near the wall is $\Delta r^+\approx 0.78$ and
$(R \Delta \theta)^+ \approx 4.9$, where the radius $r^+$ and the
arc length at the wall $(R \Delta \theta)^+$ are normalized by wall units $\nu /
U_\tau$ denoted by the superscript $+$.  Near the center of the
pipe, the grid width is $\Delta^+\approx 3.1$.  The grid
spacing in the streamwise direction is a constant $\Delta z^+=6.25$ throughout the domain.
Further details can be found in Duggleby
et al.\cite{duggleby_JOT}

The flow is driven by a constant mean pressure gradient to keep
$\mathrm{Re}_\tau$ constant.  The spanwise wall
oscillation results in a mean flow rate increase, effectively changing
the Reynolds number based upon mean velocity ($\mathrm{Re}_m$) while
keeping $\mathrm{Re}_\tau$ constant.  This keeps the dominant structures of the flow
similar, as they are affected primarily by the inner layer wall shear
stress.\cite{zhou_dongmei}  The oscillations were started on a fully
turbulent pipe at $\mathrm{Re}_\tau=150$, and to avoid transient effects, data were not taken until the mean flow rate had
settled at its new average value over a time interval of $1000 t^+$.
 
In the Karhunen-Lo\`{e}ve (KL) procedure, the eigenfunctions of the
two-point velocity correlation tensor, defined by

\begin{eqnarray}
\label{KL}
\int_0^L \int_0^{2 \pi} \int_0^R K(\mathbf{x}, \mathbf{x}') \mathbf{\Phi}
(\mathbf{x}') r'dr' d \theta' dz' = \lambda \mathbf{\Phi}(\mathbf{x}), \\
\label{two_point}
K(\mathbf{x}, \mathbf{x'})=\langle \mathbf{u} (\mathbf{x}) \otimes \mathbf{u}
(\mathbf{x}') \rangle,
\end{eqnarray}
are obtained, where
$\mathbf{x}=(r,\theta,z)$ is the position vector, $\mathbf{\Phi} (\mathbf{x})$
is the eigenfunction with associated eigenvalue $\lambda$,
$K(\mathbf{x}, \mathbf{x}')$ is the kernel, and $\otimes$ denotes an
outer product.  In order to focus on the
turbulent structures, the kernel is built using fluctuating velocities
 $\mathbf{u}=\mathbf{U}-\overline{\mathbf{U}}$.  The mean
velocity, $\overline{\mathbf{U}}$, is found by averaging over all 
$\theta$, $z$, and time.  The angle brackets in equation \ref{two_point} represent the time
average using an 
evenly spaced time interval over a total time period sufficient to
sample the turbulent attractor.  In this study, the
flow field was sampled every $8 t^+$ for a total time of $16800 t^+$.

Since the azimuthal and streamwise
directions are periodic, the kernel in the azimuthal and streamwise
direction is only a function of the distance between $\mathbf{x}$ and
$\mathbf{x}'$ in those respective directions.  Therefore the kernel can be rewritten as
\begin{eqnarray}
\label{kernel}
K(r, \theta,z,r',\theta',z')& = &K(r,r', \theta-\theta', z-z') \nonumber \\
& = & \mathcal{K} (m,n; r,r')  e^{i n\theta} e^{i 2  \pi m z / L}
\end{eqnarray}
 with azimuthal and streamwise wavenumbers $n$ and $m$
respectively and the remaining two-point correlation in the radial
direction $\mathcal{K}(m,n; r,r')$.  It can be shown that in this form, the Fourier series is
 the resulting KL function in the streamwise and azimuthal
direction. \cite{Holmes}  The resulting eigenfunction then takes the form
\begin{equation}
\label{psi}
\mathbf{\Phi}(r,\theta,z)=\mathbf{\Psi}(m,n; r) e^{i n\theta} e^{i 2 \pi m z / L}.
\end{equation}
Making use of this result, and noting that the 
two-point correlation in a periodic direction is simply the Fourier
transform of the velocities, the azimuthal and streamwise
contributions to the eigenfunctions are extracted {\it a priori} by
taking the Fourier transform of the velocities  $\mathbf{u} (r,\theta,z) = \sum_{m,n=0}^{\infty}
\hat{\mathbf{u}} (m,n; r) e^{i n\theta} e^{i 2  \pi m z / L}$ and forming the
remaining kernel $\mathcal{K}(m,n; r,r')$ for each wavenumber pair $n$
and $m$.
  The eigenfunction problem, with the orthogonality of
the Fourier series taken into account, is 
\begin{eqnarray}
\label{reduced_inner}
\int_0^R \mathcal{K}(m,n; r,r')\mathbf{\Psi}^\star(r') r'
dr'=\lambda_{mn}\mathbf{\Psi} (m,n; r), \\
\label{reduced}
\mathcal{K} (m,n; r,r') = \langle \hat{\mathbf{u}} (m,n; r) \otimes
\hat{\mathbf{u}}^\star (m,n; r') \rangle. 
\end{eqnarray}
where the $\star$ denotes the complex conjugate since the function is
 now complex, and the weighting function $r'$ is
 present because the inner product is evaluated in polar-cylindrical
 coordinates.  The final form is still Hermitian just as it was in
 equation \ref{KL}; the 
 discrete form of equation \ref{reduced_inner} is kept Hermitian by
 splitting the integrating weight and solving the related eigenvalue problem
\begin{eqnarray}
\label{hermitian}
 \left[ \sqrt{r_p} \mathbf{\mathcal{K}}_{ps}(m,n; r_p,r_s) \sqrt{r_s} \right]
\left[\sqrt{r_s} \mathbf{\Psi}_{q}(m,n; r_s) \right]^\star \nonumber \\
 = \lambda_{mnq}
\left[ \sqrt{r_p} \mathbf{\Psi}_{q} (m,n,; r_p) \right],
\end{eqnarray}
where $\mathbf{\mathcal{K}}_{ps}(m,n; r_p,r_s)$ is the discretization of
$\mathcal{K} (m,n; r,r')$ using a $Q$ point quadrature to evaluate
equation \ref{reduced_inner} with
$p,s=1,2,...,Q$.  Because the kernel is built with the two-point
correlation between all three coordinate velocities, its solution has
$3Q$ complex eigenfunctions $\mathbf{\Psi}_{q}$ and
corresponding eigenvalues, listed in decreasing order
$\lambda_{mnq}>\lambda_{mn(q+1)}$  for a given $m$ and $n$, with
quantum number $q=1,2,...,3Q$.  It is
noted that equation \ref{hermitian} is only valid for a trapezoidal
integration scheme with evenly spaced grid points (which was used in
this study), a different quadrature with weight $w(r)$ can be
incorporated in a similar fashion, keeping the final matrix Hermitian.

The eigenfunctions$\mathbf{\Psi}_q(m,n; r)$ hold certain
properties.  Firstly, they are normalized with inner product
of unit length 
$\int_0^R \mathbf{\Psi}_q,\mathbf{\Psi}_{q'} r dr=\delta_{qq'}$,
where $\delta$ is
the Kronecker delta.  Secondly, since the eigenfunctions represent
a flow field,
\begin{equation}
\label{eigen_flow}
\mathbf{\Psi}_q(m,n; r)=
\left(\Psi^r_{q} (m,n; r), \Psi^{\theta}_{q} (m,n; r), \Psi^z_{q} (m,n; r) \right)^T
\end{equation}
with radial, azimuthal, and streamwise components $\Psi^r_{q}
 (m,n; r)$, $\Psi^{\theta}_{q} (m,n;r)$, and $\Psi^z_{q} (m,n; r)$
 respectively, they hold the properties of the flowfield such as
 boundary conditions (no slip) and continuity
\begin{equation}
\label{cont_eigen}
\frac{1}{r}\frac{d}{dr}(r \Psi^r_{q} (m,n; r))+\frac{i n}{r} \Psi^{\theta}_{q}(m,n;
 r)+\frac{i 2 \pi m}{L}
 \Psi^z_{q}(m,n; r) = 0.
\end{equation}
Thirdly, the eigenvalues represent the average energy of the flow
  contained in the eigenfunction $\mathbf{\Psi}_{q}(m,n; r)$

\begin{equation}
\label{energy}
\lambda_{mnq}=\langle |\mathbf{u}(r, \theta, z), \mathbf{\Psi}_{q}(m,n; r)e^{i n
  \theta}e^{i 2 \pi m z / L}|^2 \rangle,
\end{equation}
which is why it is necessary that the discrete matrix in equation
\ref{hermitian} must be Hermitian to get real and positive eigenvalues.
These three properties allow the eigenfunctions to be tested for their validity.

In summary, the KL procedure yields an orthogonal set of
basis functions (modes) that are the most energetically efficient expansion of the flow
field.  By studying the subset that includes the largest energy modes,
insight is gained as this subset forms a low dimensional model of the
given flow.  Examining the structure, dynamics, and
interactions of this low dimensional model yields important
information that we use to build a better understanding of the
dynamics of the entire system. 

%********* RESULTS**********

\section{RESULTS}
Spanwise wall oscillation results in four main effects on the flow,
its structures, and its dynamics. They are:

\begin{enumerate}
\item An increase in flow rate and a shifting away from the wall of
  the root-mean-squared velocities and Reynolds stress peaks.
\item A reduction in the dimension of the chaotic attractor describing
  the turbulence.
\item An increase in energy of the propagating modes responsible for
  carrying energy away from the wall to the upper region, while the
  rest of the propagating modes exhibit a decrease in energy.
\item An increase of the advection speed of the traveling wave
  packet as determined by a normal speed locus.
\end{enumerate}

\begin{figure}
\includegraphics[width=3 in]{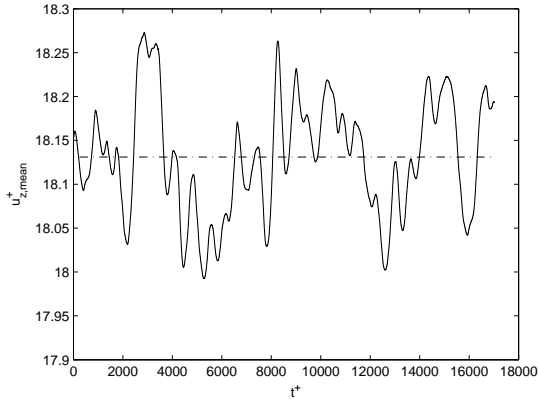}
\caption{Mean velocity fluctuations (solid) and the
  average velocity (dash-dot) for the oscillated case versus time
  ($t^+$).  The fluctuations are consistent with turbulent flow, and the mean velocity is 26.9\% greater than
  in the non-oscillated pipe.}
\label{mean_time}
\end{figure}

\begin{figure}[h]
%\centering
%\begin{tabular}{cc}
%\begin{minipage}{3 in}
\includegraphics[width=3 in]{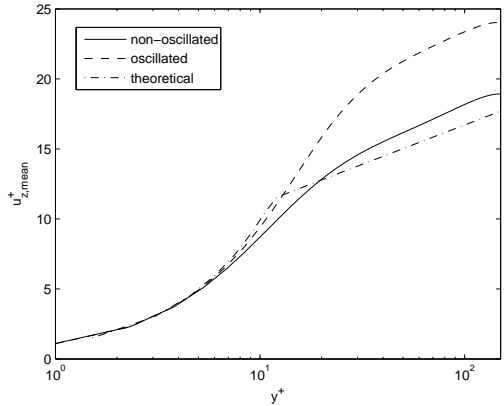}
\caption{Mean velocity profile for a
  non-oscillated (solid) and
  oscillated (dashed) turbulent pipe flow versus $y^+$.  Theoretical (dash-dot) includes the
  sublayer ($u^+=y^+$) and the log layer ($u^+=\log (y^+)/0.41+5.5$).
  The mean profile shows a log layer, but overshoots the theoretical
  value as expected for pipe flow until a much higher Reynolds number.}
\label{log_mean_profile}
\end{figure}

\begin{figure}
\includegraphics[width=3in]{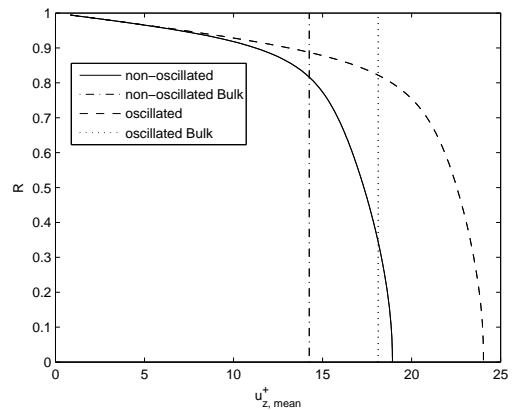}
\caption{Mean velocity profile for non-oscillated (solid) and
  oscillated (dashed) pipes with their respective bulk velocities (dash-dot and
  dots) versus radius show an increase in bulk velocity
  of 26.9\%.}
\label{mean_profile}
%\end{minipage} \\
%\end{tabular}
 \end{figure}

 First, surface oscillation has a major effect on the turbulent statistics.
The spanwise wall oscillation of amplitude $A^+=20$ and period
$T^+=50$ resulted in a flow rate increase of 26.9\%, shown in Figure
\ref{mean_time}.  This combination of amplitude and period was chosen because it provides
the largest amount of drag reduction while keeping the flow
turbulent. Numerical simulations with larger oscillations completely relaminarize the flow.  The comparison of the mean profiles
in Figures \ref{log_mean_profile} and \ref{mean_profile} show the
higher velocity in the outer region, with the inner region remaining
the same, as expected by keeping a constant mean pressure gradient across
the pipe. 

%\end{minipage}
%& 
%\begin{minipage}{3 in}
%\addtocounter{figure}{-1}
%\addtocounter{subfigure}{1}

\begin{figure}[h]
%\begin{tabular}{cc}
%\begin{minipage}{2.5in}
\includegraphics[width=3 in]{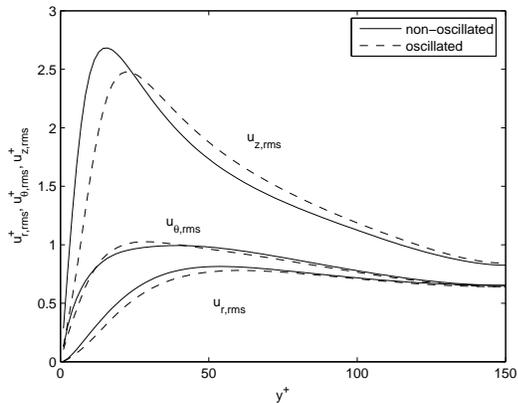}
\caption{Root-mean-squared velocity fluctuations for
  the non-oscillated (solid) and oscillated (dashed) case versus
  $y^+$.  The oscillated case shows a
  shift away from the wall, except for the azimuthal rms which
  captures the Stokes layer imposed by the wall oscillation.}
\label{rms}
\end{figure}

\begin{figure}
\includegraphics[width=3in]{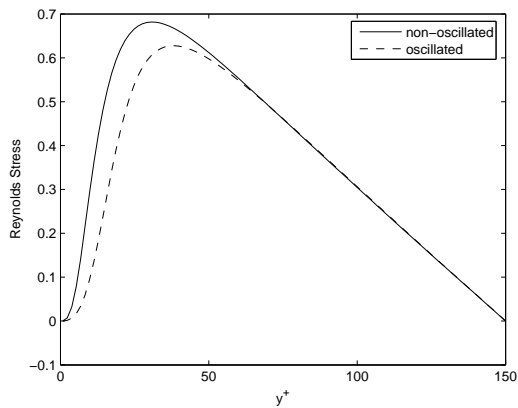}
\caption{Reynolds Stress $\overline{u_r u_z}$ versus $y^+$ for
  non-oscillated (solid) and oscillated (dashed) cases.  Similar to the rms velocities,
  the Reynolds stress shows a shift away from the wall from $y^+=31$
  to $y^+=38$. }
%\end{minipage} \\
%\end{tabular}
\label{reynolds}
 \end{figure}

The comparison of the root-mean-square (rms) velocity fluctuation profiles
and the Reynolds stress profile in Figures
\ref{rms} and \ref{reynolds} show that the streamwise fluctuations
decrease in intensity by 7.5\%
from 2.68 to 2.48.  Also, the change in peak location from $y^+=16$ to
$y^+=22$ away from the wall has the same trend as the maximum Reynolds stress  $\overline{u_r
  u_z}$, where $y^+=(R-r)U_\tau / \nu$ is the distance from the wall using
normalized wall units ($\nu / U_\tau$).  The
azimuthal fluctuating velocities show a slightly greater magnitude peak of
1.03 closer to the wall at $y^+=29$ versus 0.99 at $y^+=40$ for
the non-oscillated pipe.  The radial fluctuations remain almost
unchanged, showing only a slight decrease from the wall
through the log layer ($y^+ \approx 100$), resulting in the peak
shifting from 0.81 at $y^+=55$ to 0.78 at $y^+=61$.  The Reynolds
stress also shows a reduction in strength and a
shift away from the wall.  The peak changes from 0.68 at $y^+=31$ to
0.63 at $y^+=38$.  Thus, a major difference between the two flow
cases, in addition to the expected flow rate increase, is the shift of
the rms velocity and Reynolds stress peaks away from the wall.

%\end{minipage}
%& 
%\begin{minipage}{2.5 in}
%\addtocounter{figure}{-1}
%\addtocounter{subfigure}{1}

\begin{figure}
\includegraphics[width=3 in]{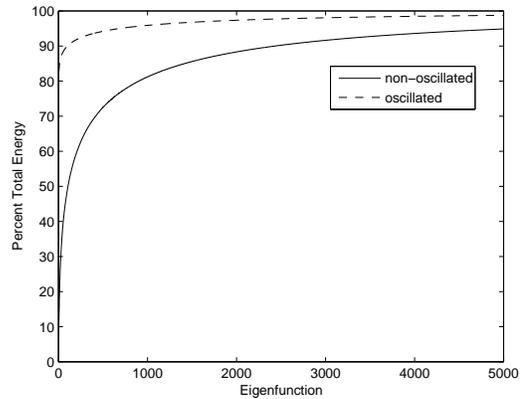}
\caption{Comparison of the running total energy retained in the KL expansion for
  the non-oscillated (solid) and oscillated (dashed) cases.  The 90\%
  crossover point is 2453 and 102 respectively.  This shows a drastic
  reduction in the dimension of the chaotic attractor.}
\label{energyEigen}
\end{figure}

The second major effect can be found by examining the size of the chaotic
attractor describing the turbulence. The eigenvalues of the KL decomposition represent the energy of
each eigenfunction.  By ordering the eigenvalues from largest to
smallest, the number of eigenfunctions needed to capture a given
percentage of energy of the flow
is minimized.  Table \ref{top25} shows the 25 most energetic eigenfunctions, and
Figure \ref{energyEigen} shows the running total of energy versus mode
number.
90\% of the energy is reached with $D_{KL}=102$ compared to $D_{KL}=2453$ for the
non-oscillated case.  This mark, known as the Karhunen-Lo\`{e}ve
dimension, is a measure of the intrinsic dimension of
the chaotic attractor of turbulence as discussed by Sirovich
\cite{sirovich_chaos, zhou_sirovich}.  By oscillating the
wall our results show that the size of the attractor is reduced, and
the system is less chaotic.

The third major effect is found by examining 
the energy of the eigenfunctions.  The 25 most energetic are listed for each case in Table
\ref{eigenList}.  The top ten modes with the largest change in energy
are shown in Table \ref{energy_change}.  Firstly, the order of the
eigenfunctions remain relatively unchanged, with a few
notable differences.  The (0,0,1) and (0,0,3) shear modes represent the
Stokes flow as seen in Figures \ref{001real} and \ref{003real}, and the
(0,0,2) mode, shown in Figure \ref{002real}, represents the changing
of the mean flow rate, similar to the non-oscillated (0,0,1) mode.
The (1,2,1) mode shows a large increase in energy.
Also of note is the reduction in energy of the (3,1,1)
and (4,1,1) modes, but their structure remains virtually unchanged.  

\begin{table}
\caption{The 25 most energetic modes; $m$ is the streamwise
  wavenumber, $n$ is the spanwise wavenumber, and $q$ is the eigenvalue
  quantum number.}
\label{eigenList}
\centering
\begin{tabular}{lcccclcccccc}
\hline
\hline
& \multicolumn{5}{c}{Non-oscillated}  & &
\multicolumn{5}{c}{Oscillated}  \\
Index & $m$ & $n$ & $q$ &  Energy & \% Total & & $m$ & $n$ & $q$ & Energy & \%
Total \\
\hline
1 & 0 & 6 & 1 & 1.61 & 2.42\%  & & 0 & 0 & 1 & 216 & 68.77\% \\
2 & 0 & 5 & 1 & 1.48 & 2.22\%  & & 0 & 0 & 2 & 34.7 & 11.07\%\\
3 & 0 & 3 & 1 & 1.45 & 2.17\%  & & 1 & 2 & 1 & 2.47 & 0.79\%\\
4 & 0 & 4 & 1 & 1.29 & 1.93\%  & & 0 & 3 & 1 & 2.30 & 0.73\%\\
5 & 0 & 2 & 1 & 1.26 & 1.88\%  & & 0 & 1 & 1 & 2.27 & 0.73\%\\
6 & 1 & 5 & 1 & 0.936 & 1.40\% & & 0 & 2 & 1 & 2.20 & 0.70\%\\
7 & 1 & 6 & 1 & 0.917 & 1.37\% & & 0 & 4 & 1 & 1.49 & 0.48\%\\
8 & 1 & 3 & 1 & 0.902 & 1.35\% & & 1 & 3 & 1 & 1.09 & 0.35\%\\
9 & 1 & 4 & 1 & 0.822 & 1.23\% & & 0 & 5 & 1 & 1.04 & 0.33\%\\
10 & 0 & 1 & 1 & 0.805 & 1.20\% & & 1 & 1 & 1 & 0.953 & 0.30\% \\
11 & 1 & 7 & 1 & 0.763 & 1.14\% & & 1 & 4 & 1 & 0.772 & 0.25\%\\
12 & 1 & 2 & 1 & 0.683 & 1.02\% & & 0 & 6 & 1 & 0.653 & 0.21\%\\
13 & 0 & 7 & 1 & 0.646 & 0.97\% & & 0 & 0 & 3 & 0.616 & 0.20\%\\
14 & 2 & 4 & 1 & 0.618 & 0.92\% & & 1 & 5 & 1 & 0.582 & 0.19\%\\
15 & 0 & 8 & 1 & 0.601 & 0.90\% & & 1 & 6 & 1 & 0.542 & 0.17\%\\
16 & 2 & 5 & 1 & 0.580 & 0.87\% & & 2 & 3 & 1 & 0.490 & 0.16\%\\
17 & 1 & 1 & 1 & 0.567 & 0.85\% & & 0 & 1 & 2 & 0.482 & 0.15\%\\
18 & 2 & 7 & 1 & 0.524 & 0.78\% & & 2 & 4 & 1 & 0.468 & 0.15\%\\
19 & 1 & 8 & 1 & 0.483 & 0.72\% & & 2 & 5 & 1 & 0.444 & 0.14\%\\
20 & 2 & 6 & 1 & 0.476 & 0.71\% & & 0 & 7 & 1 & 0.407 & 0.13\%\\
21 & 2 & 3 & 1 & 0.454 & 0.68\% & & 1 & 7 & 1 & 0.373 & 0.12\%\\
22 & 2 & 2 & 1 & 0.421 & 0.63\% & & 2 & 6 & 1 & 0.346 & 0.11\%\\
23 & 2 & 8 & 1 & 0.375 & 0.56\% & & 2 & 2 & 1 & 0.337 & 0.11\%\\
24 & 1 & 9 & 1 & 0.358 & 0.54\% & & 3 & 5 & 1 & 0.317 & 0.10\%\\
25 & 3 & 4 & 1 & 0.354 & 0.53\% & & 3 & 3 & 1 & 0.290 & 0.09\%\\
\hline
\hline
\end{tabular}
\label{top25}
\end{table}

\begin{table}
\caption{Ranking of eigenfunctions by energy change between the
  non-oscillated and the oscillated cases.  $m$ is the streamwise
  wavenumber, $n$ is the spanwise wavenumber, and $q$ is the eigenvalue
  quantum number.}
\label{energy_change}
\centering
\begin{tabular}{llcccllccc}
\toprule
 &  \multicolumn{4}{c}{Increase} & & \multicolumn{4}{c}{Decrease} \\
Rank & $\Delta \lambda_{\mathbf{k}}$ & $m$ & $n$ & $q$ & & $\Delta \lambda_{\mathbf{k}}$ & $m$ & $n$ & $q$ \\
\colrule
1 & 215 & 0 & 0 & 1 & &-0.962 & 0 & 6 & 1 \\
2 & 34.5 & 0 & 0 & 2 & &-0.442 & 0 & 5 & 1 \\
3 & 1.79 & 1 & 2 & 1 & &-0.390 & 1 & 7 & 1 \\
4 & 1.47 & 0 & 1 & 1 & &-0.382 & 0 & 8 & 1 \\
5 & 0.95 & 0 & 2 & 1 & &-0.374 & 1 & 6 & 1 \\
6 & 0.85 & 1 & 3 & 1 & &-0.353 & 1 & 5 & 1 \\
7 & 0.49 & 0 & 0 & 3 & &-0.256 & 1 & 8 & 1 \\
8 & 0.37 & 1 & 1 & 1 & &-0.245 & 2 & 7 & 1 \\
9 & 0.26 & 0 & 1 & 2 & &-0.239 & 0 & 7 & 1 \\
10 & 0.20 & 0 & 4 & 1 & & -0.220 & 2 & 8 & 1 \\
\botrule
\end{tabular}
\end{table}

\renewcommand{\thefigure}{\arabic{figure}\alph{subfigure}}
\setcounter{subfigure}{1}
\begin{figure}
\includegraphics[width=3 in]{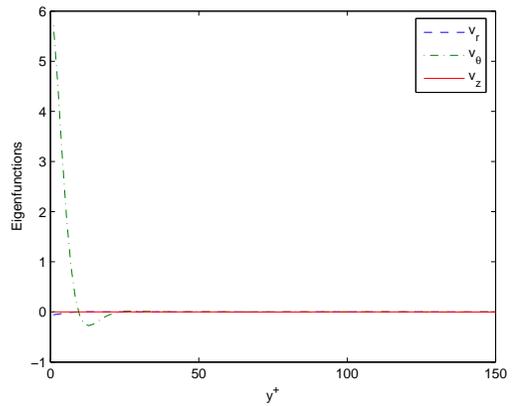}
\caption{(0,0,1) eigenfunction capturing the imposed velocity
  oscillations.  The radial velocity is zero except near the wall,
  where a slight non-zero component is evident, induced by the spanwise oscillation.}
\label{001real}
\end{figure}

\addtocounter{figure}{-1}
\addtocounter{subfigure}{1}
\begin{figure}
\includegraphics[width=3 in]{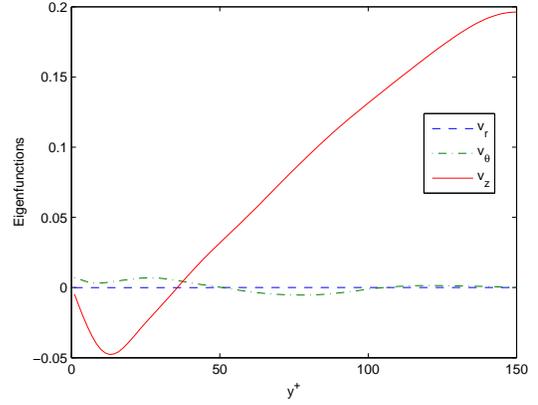}
\caption{(0,0,2) eigenfunction, similar in structure to the
  non-oscillated (0,0,1).}
\label{002real}
\end{figure}

\addtocounter{figure}{-1}
\addtocounter{subfigure}{1}
\begin{figure}
\includegraphics[width=3 in]{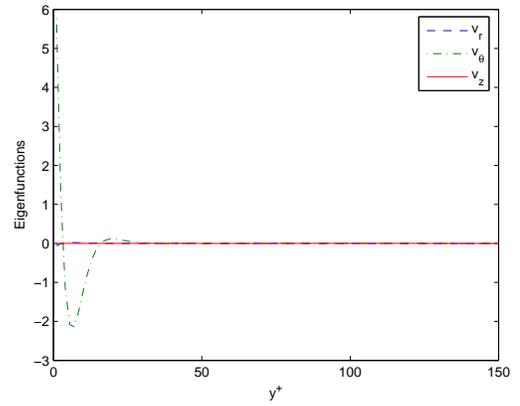}
\caption{(0,0,3) eigenfunction capturing the secondary flow in the
  Stokes layer.  This imposed Stokes flow created by the oscillation dominates the near wall
  region of the oscillated pipe.}
\label{003real}
\end{figure}
\renewcommand{\thefigure}{\arabic{figure}}

In examining the energy content of the structure subclasses as a
whole, a trend is discovered, shown in Table \ref{subclassShift}.
Each of these subclasses, reported in Duggleby et al.,
\cite{duggleby_JOT} were found to 
have similar qualitative coherent vorticity structure associated with
their streamwise and azimuthal wavenumber.  We use ``coherent vorticity''
to refer to the imaginary part of the eigenvalues of the
strain rate tensor $\partial u_i / \partial x_j$,
following the work of Chong et al.. \cite{chong}  Based upon the
qualitative structure, the propagating or traveling waves, described
by non-zero azimuthal wavenumber and nearly constant phase speed, were found
to have four subclasses: the wall, the lift, the asymmetric, and the
ring modes.    

The wall modes are found when the
spanwise wavenumber is larger than the streamwise wavenumber.  They possess
a 
qualitative structure having
coherent vortex cores near the wall, and have their energy decreased
by 20.4\% with wall oscillation.  Likewise, the
ring modes, which are found for
non-zero streamwise wavenumber and zero spanwise wavenumber with rings
of coherent vorticity, have their energy decreased by 5.2\% with wall oscillation.
The asymmetric modes have non-zero streamwise wavenumber and spanwise
wavenumber $n=1$, which
allows them to break azimuthal
symmetry.  These modes
undergo a decrease in energy of 2.3\% with spanwise wall oscillation.

Conversely to the decrease in energy found in the other three
propagating modes, the lift modes increase in energy by 1.5\% with
spanwise wall oscillation.  These modes are found with a streamwise
wavenumber that is greater than the
spanwise wavenumber, and they display coherent vortex structures that starts
near the wall and lifts away from the
the wall to the upper region.  Combined, the four propagating
subclasses modes lose
9.02\% of their energy, whereas the non-propagating modes (the modes with zero azimuthal wavenumber)
gain 2043\%. 

\begin{figure}[h]
\includegraphics[width=3.5 in]{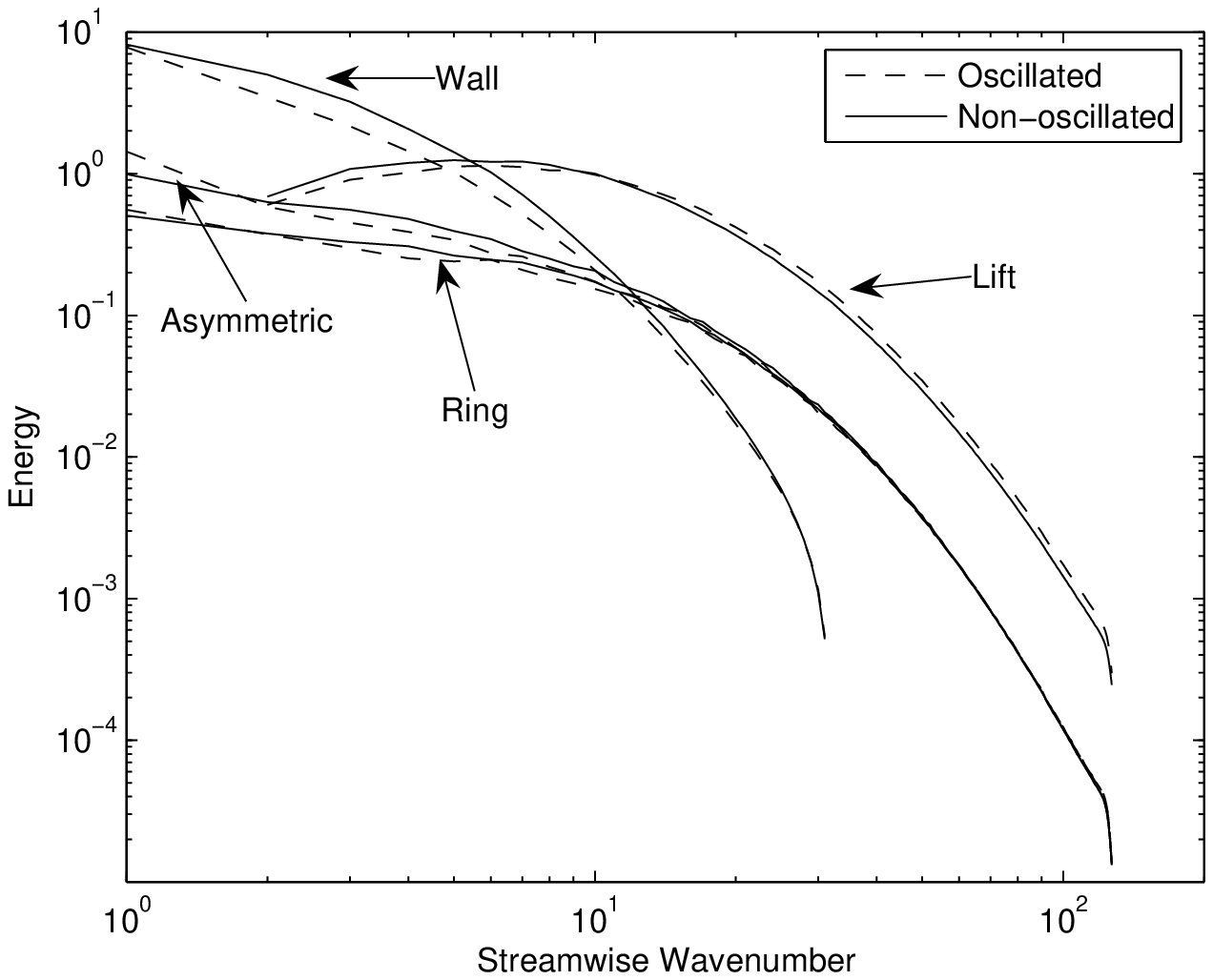}
\caption{Comparison of energy spectra for non-oscillated (solid) and
  oscillated (dashed) flows for the propagating mode subclasses.  All
  propagating subclasses in the oscillated case decrease with respect to the
  non-oscillated case, except for the lift mode, which increases slightly.}
\label{distribE}
\end{figure}

 Thus, the third major result is that the
energy of the propagating wall, ring, and asymmetric modes decrease while the energy of lift
mode 
increases slightly.  Following the work of Sirovich et
al. \cite{sirovich2} this shows that energy transfer from
the streamwise rolls to the traveling waves is reduced, and any energy
that is transferred is quickly moved away from the wall to the outer
region (lift modes).  The energy spectra showing
the change of energy by subclasses is shown in Figure \ref{distribE}.

\begin{table}
\caption{Energy comparison of turbulent pipe flow structure subclasses
  between non-oscillated and oscillated pipes. $m$ is the streamwise
wavenumber, and $n$ is the azimuthal (spanwise) wavenumber.  All the
  propagating modes decrease in energy, except the lift modes. }
\label{subclassShift}
\begin{tabular}{lll}
\hline \hline
\bf Structure & \multicolumn{2}{c}{\bf Energy} \\
&  Non-oscillated & Oscillated \\
\hline
Propagating Modes $(m>0)$ &  53.48  & 48.86 \\
\hspace{2em}(a) Wall $(n>m)$ & 23.5 & 18.7\\
\hspace{2em}(b) Lift $(m \geq n)$ & 19.8 & 20.1 \\
\hspace{2em}(c) Asymmetric $(n=1)$ & 6.07 & 5.93\\
\hspace{2em}(d) Ring $(n=0)$ & 4.41 & 4.18 \\
& & \\
Non-propagating Modes $(m=0)$ & 12.97 & 265 \\
\hspace{2em}(a) Roll mode $(n>0)$& 12.3 & 13.2 \\
\hspace{2em}(b) Shear mode $(n=0)$ & 0.72 & 251.8 \\
\hline \hline
\end{tabular}
\end{table}

\begin{figure}[h]
\includegraphics[width=3.5 in]{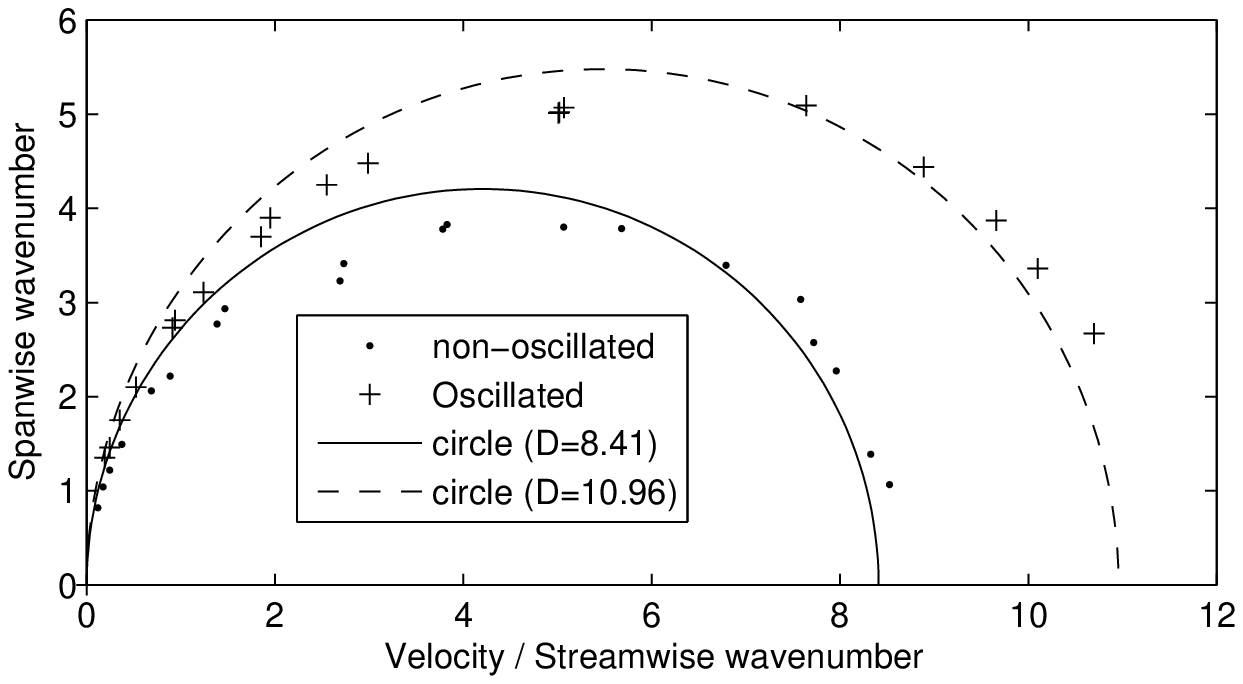}
\caption{Comparison of the normal speed locus for the oscillated ($\cdot$) and
non-oscillated ($+$) case.  The solid and dashed lines represent a circle of
diameter 8.41 and 10.96 respectively that intersect at the origin.}
\label{locus}
\end{figure}

\begin{figure}[h]
\includegraphics[width=3.5 in]{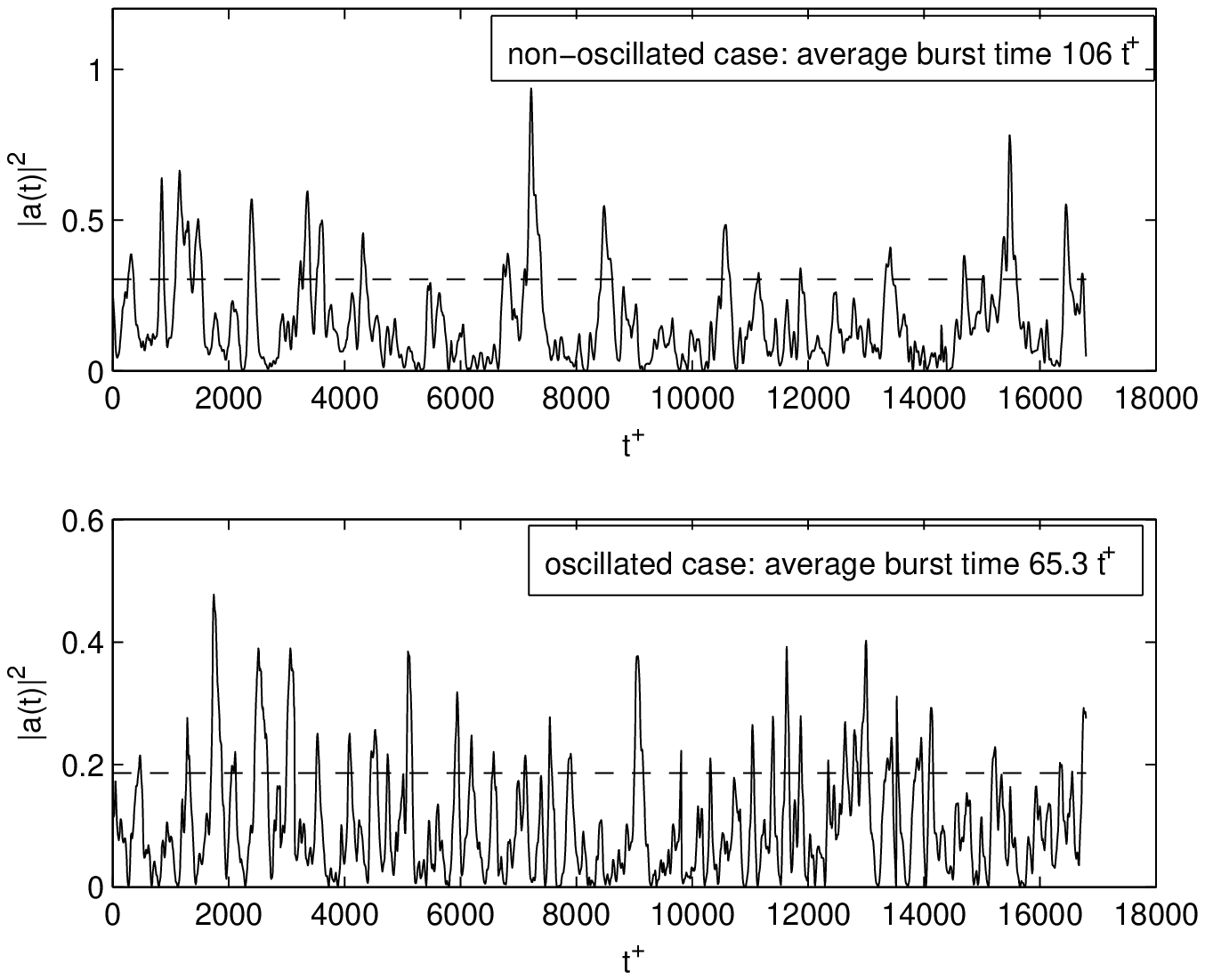}
\caption{A reduction in the burst duration of the (1,5,1) mode from $106
  t^+$ for the non-oscillated case (top)
  to $65.3 t^+$ in the oscillated case (bottom) shows a faster decay
  of the bursting energy with spanwise wall oscillation.  The burst duration is the average time of
  all events where the square of the amplitude ($|a(t)|^2$) is more
  that one standard deviation greater than the mean.  This amplitude
  level is denoted by the dashed line.}
\label{burstTime}
\end{figure}

The fourth and most important effect is that the propagating modes advect faster in the
oscillated case. The normal speed locus of the 50 most energetic
modes of both cases is shown in Figure \ref{locus}.    For this, the phase speed $\omega / \|\mathbf{ k} \|$ is plotted
in the direction $\mathbf{k} / \|\mathbf{k}\|$ with
$\mathbf{k}=(m,n)$.  A circular locus is evidence that these structures
propagate as a wave packet or envelope that travels with a constant
advection speed.  The advection speed is given by the intersection of the circle with the abscissa.  By examining the normal speed locus of the
non-oscillated and oscillated pipe flow, the wave
packet shows an increase in advection speed from $8.41 U_\tau$ to
$10.96 U_\tau$, an increase of 30\%. This is a result of the oscillating Stokes layer pushing the
structures away from the wall into a faster mean flow by creating a
dominant near wall Stokes layer where the turbulent structures cannot form.  This is
confirmed by the shifting of the rms velocities and Reynolds stresses
away from the wall as reported
earlier.  In addition to a faster advection speed, the energy of the
propagating modes decay faster, resulting in bursting events with a shorter
lifespan.  This is seen in Figure \ref{burstTime} where the average
burst duration of the (1,5,1) mode is reduced from $106 t^+$ in the
non-oscillated case to $65.3 t^+$ in
the oscillated case.  The burst duration is taken to be the average
time of all events where the square of the amplitude of the mode is
more than one standard deviation greater than the mean.

\begin{figure}
\begin{tabular}{cc}
\begin{minipage}{1.75 in}
\includegraphics[width=1.75 in]{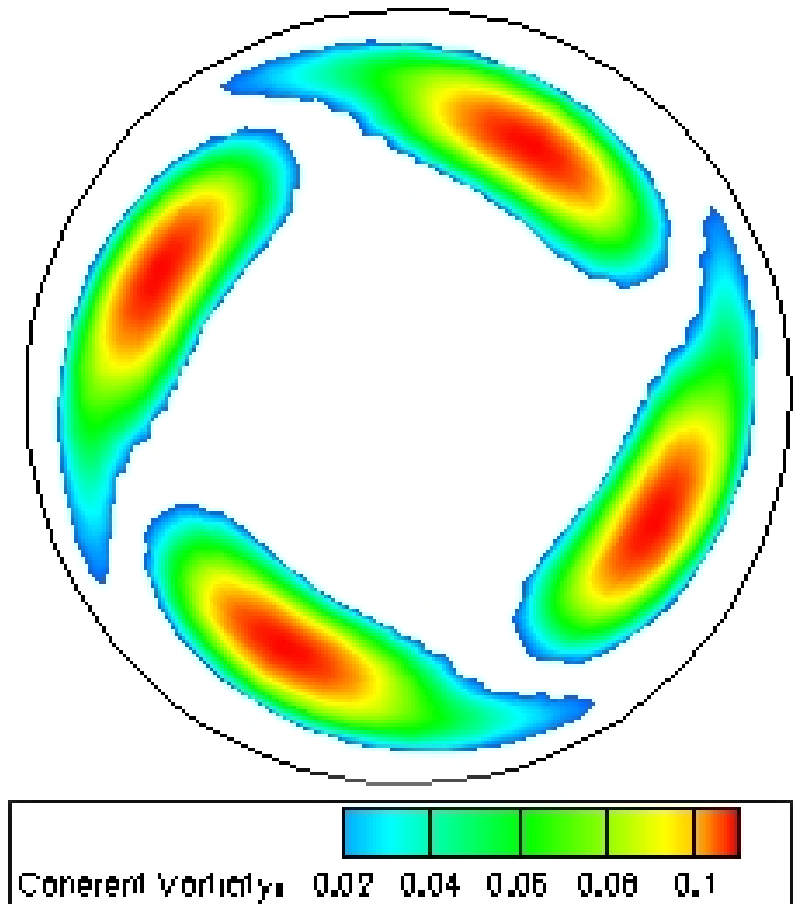}
\end{minipage}
&
\begin{minipage}{1.75 in}
\includegraphics[width=1.75 in]{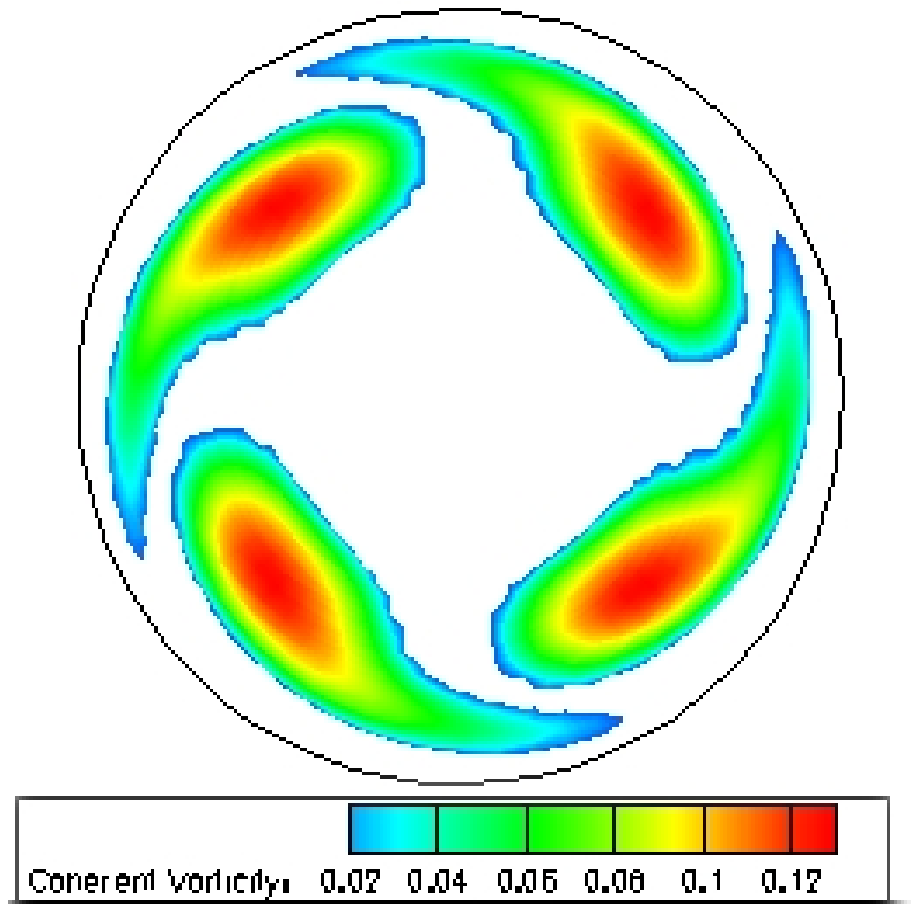}
\end{minipage}\\
\end{tabular}
\caption{Cross-section of coherent vorticity of (1,2,1) wall mode.  Left: (a)
  non-oscillated.  Right: (b) oscillated.  The vortex core shifts
  $y^+=6.8$ away from the wall.}
\label{121}
\end{figure}

\begin{figure}
\begin{tabular}{cc}
\begin{minipage}{1.75 in}
\includegraphics[width=1.75 in]{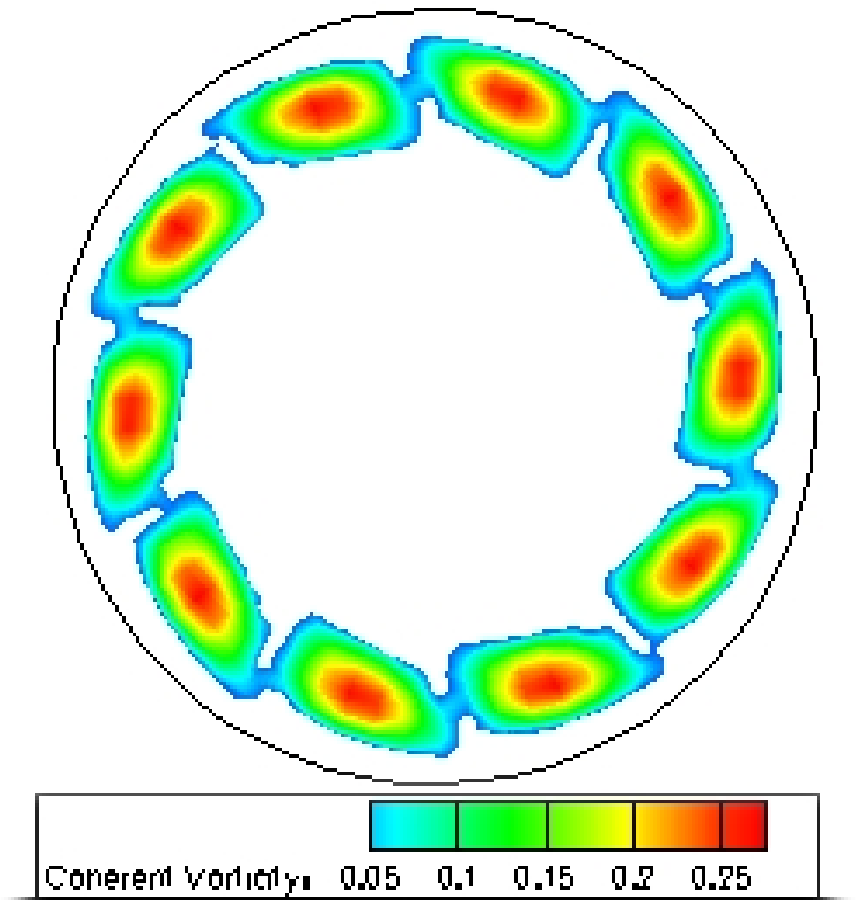}
\end{minipage}
&
\begin{minipage}{1.75 in}
\includegraphics[width=1.75 in]{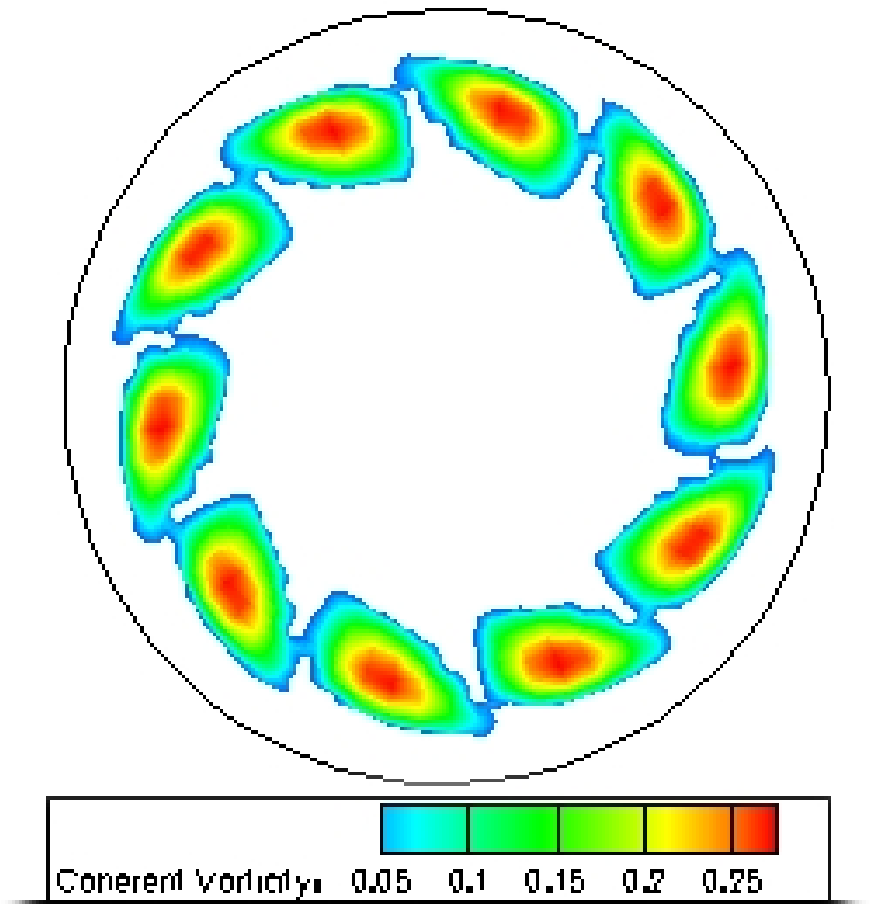}
\end{minipage}\\
\end{tabular}
\caption{Cross-section of coherent vorticity of (1,5,1) wall mode.  Left: (a)
  non-oscillated.  Right: (b) oscillated.   The vortex core shifts
  $y^+=9.9$ away from the wall.}
\label{151}
\end{figure}

\begin{figure}
\begin{tabular}{cc}
\begin{minipage}{1.75 in}
\includegraphics[width=1.75 in]{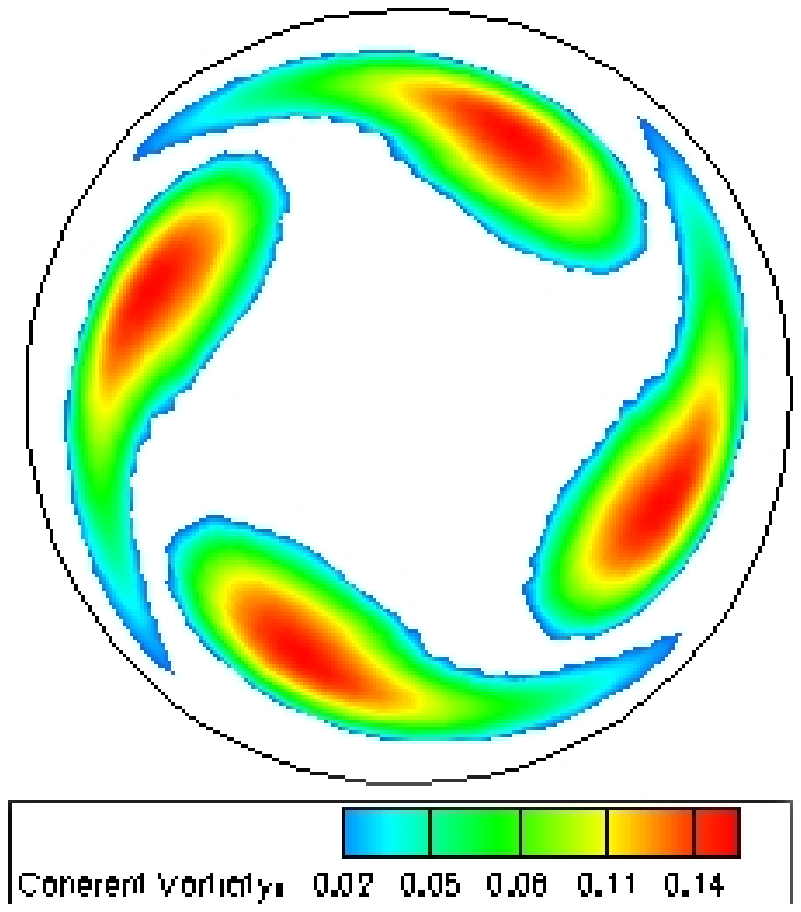}
\end{minipage}
&
\begin{minipage}{1.75 in}
\includegraphics[width=1.75 in]{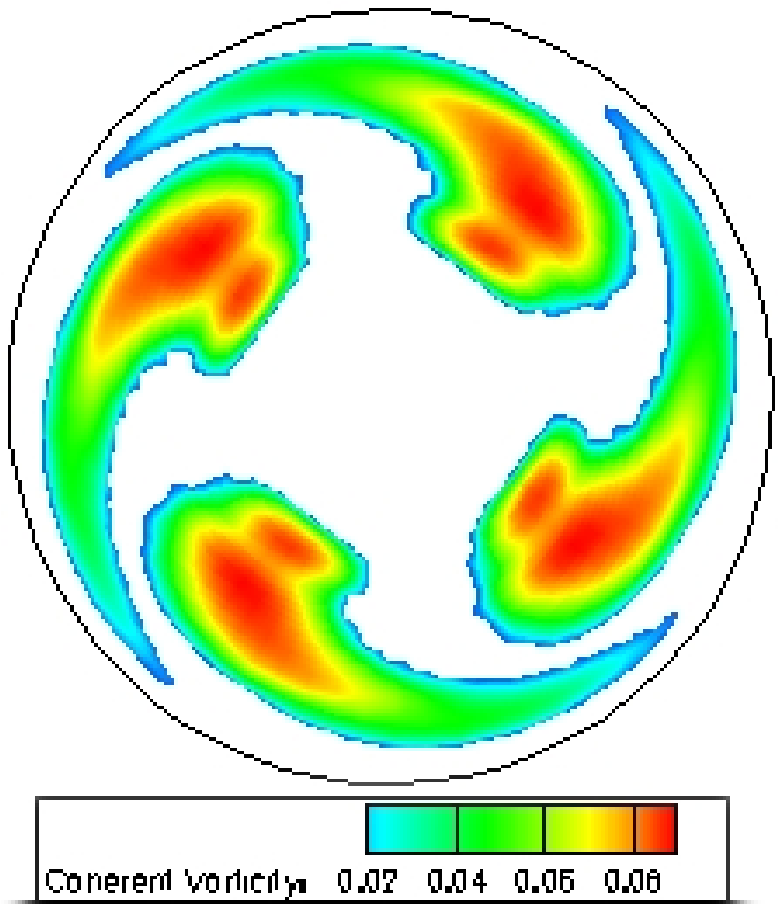}
\end{minipage}\\
\end{tabular}
\caption{Cross-section of coherent vorticity of (2,2,1) lift mode.  Left: (a)
  non-oscillated.  Right: (b) oscillated.  The vortex core shifts
  $y^+=11.0$ away from the wall.}
\label{221}
\end{figure}

\begin{figure}
\begin{tabular}{cc}
\begin{minipage}{1.75 in}
\includegraphics[width=1.75 in]{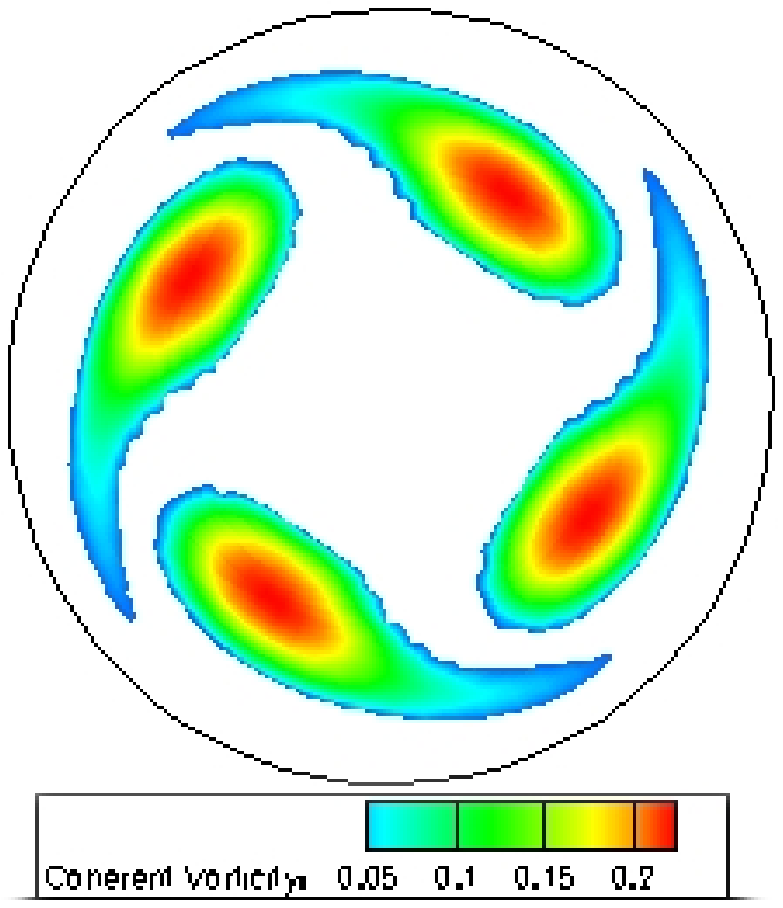}
\end{minipage}
&
\begin{minipage}{1.75 in}
\includegraphics[width=1.75 in]{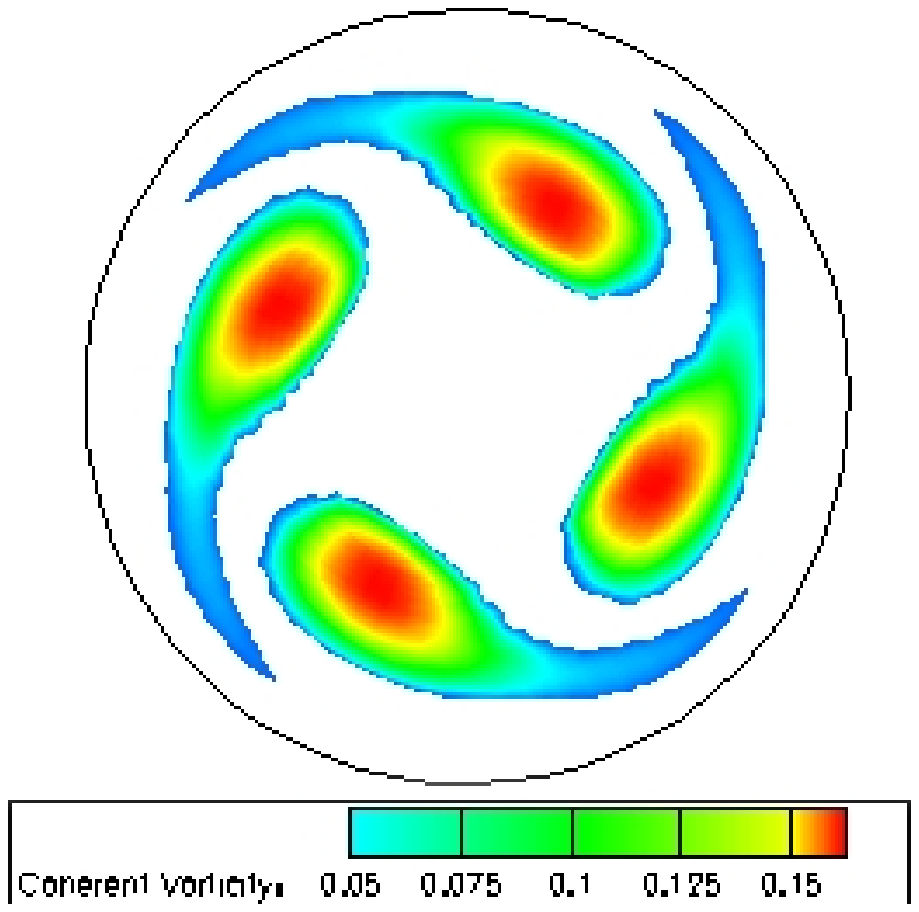}
\end{minipage}\\
\end{tabular}
\caption{Cross-section of coherent vorticity of (3,2,1) lift mode.  Left: (a)
  non-oscillated.  Right: (b) oscillated.  The vortex core shifts
  $y^+=11.2$ away from the wall.}
\label{321}
\end{figure}

\begin{figure}
\begin{tabular}{cc}
\begin{minipage}{1.75 in}
\includegraphics[width=1.75 in]{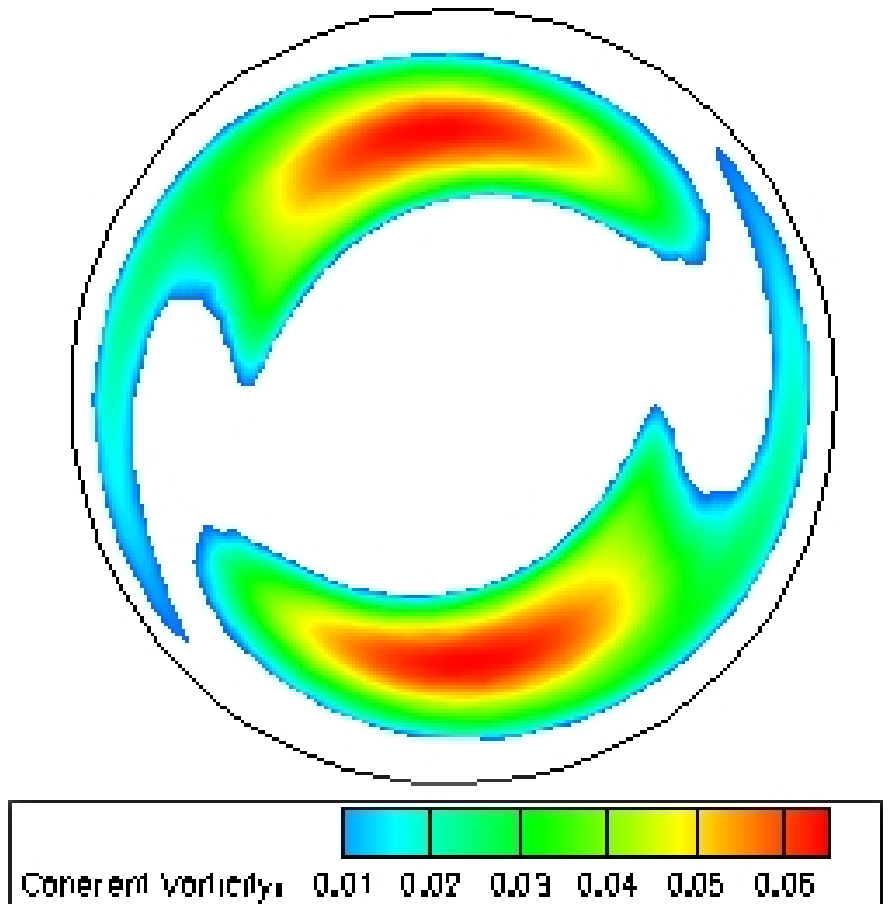}
\end{minipage}
&
\begin{minipage}{1.75 in}
\includegraphics[width=1.75 in]{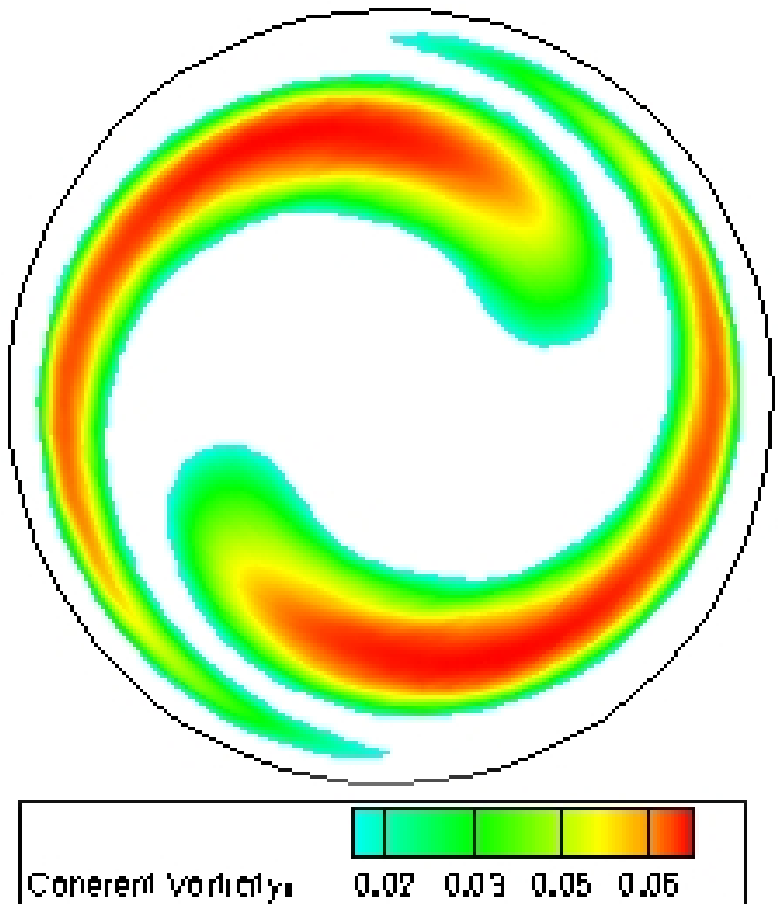}
\end{minipage}\\
\end{tabular}
\caption{Cross-section of coherent vorticity of (1,1,1) asymmetric mode.  Left: (a)
  non-oscillated.  Right: (b) oscillated.  The vortex core shifts
  $y^+=3.2$ away from the wall.}
\label{111}
\end{figure}

\begin{figure}
\begin{tabular}{cc}
\begin{minipage}{1.75 in}
\includegraphics[width=1.75 in]{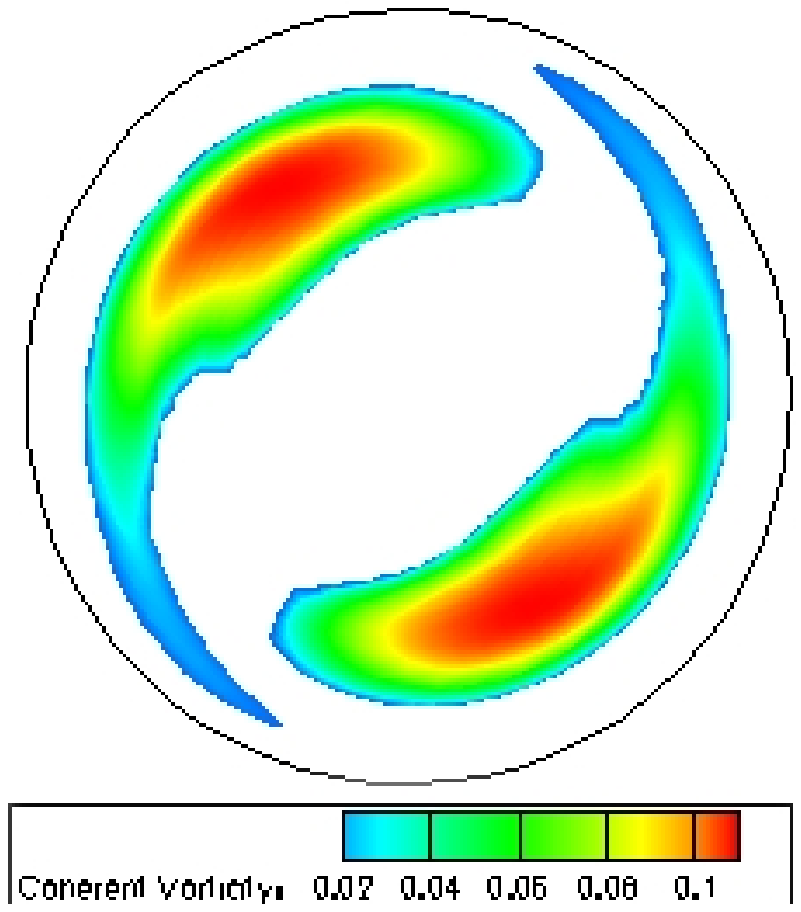}
\end{minipage}
&
\begin{minipage}{1.75 in}
\includegraphics[width=1.75 in]{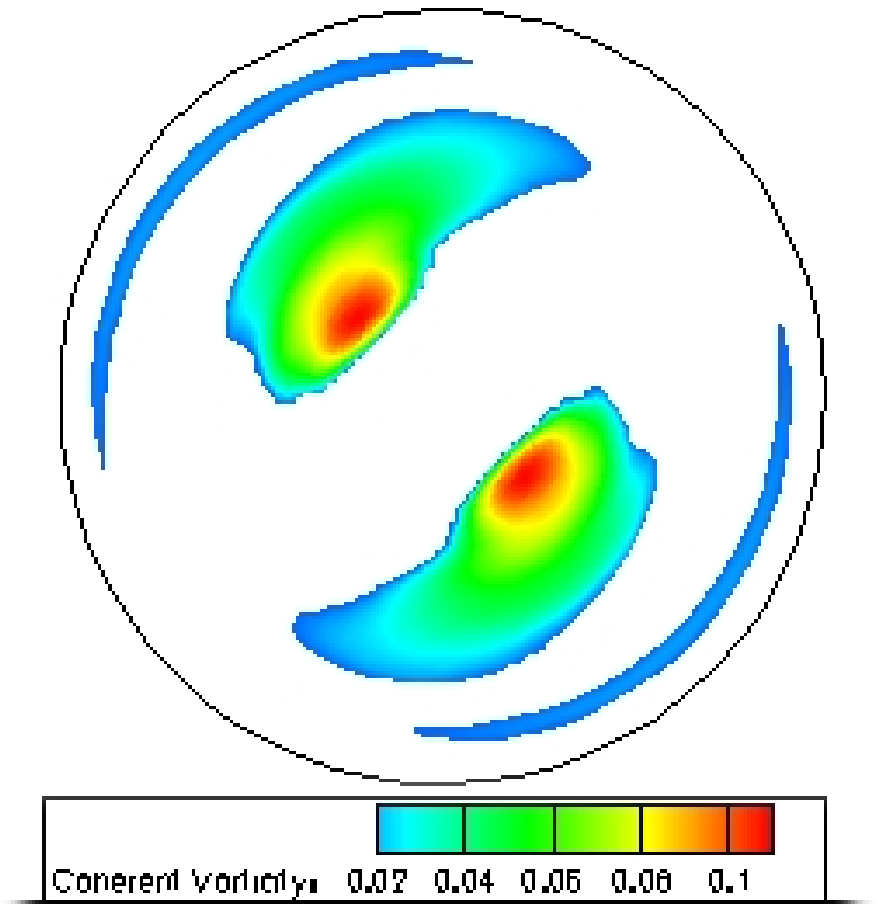}
\end{minipage}\\
\end{tabular}
\caption{Cross-section of coherent vorticity of (2,1,1) asymmetric mode.  Left: (a)
  non-oscillated.  Right: (b) oscillated.  The vortex core shifts
  $y^+=45.9$ away from the wall with a slight change in structure.}
\label{211}
\end{figure}

\begin{figure}
\includegraphics[width=3.5 in]{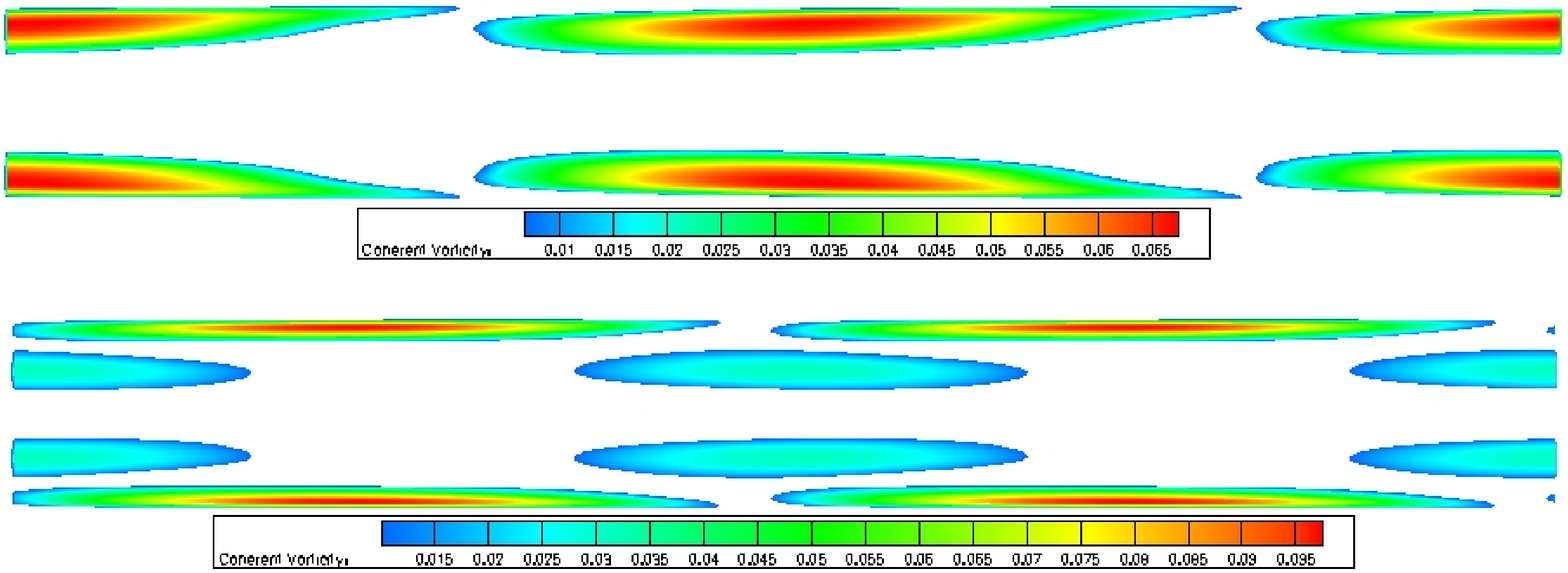}
\caption{Cross-section along the $r$-$z$ plane of coherent vorticity of (1,0,1) ring mode.  Top: (a)
  non-oscillated.  Bottom: (b) oscillated.  The vortex core shifts
  $y^+=13$ towards the wall with significant changes in structure.}
\label{101}
\end{figure}

\begin{figure}
\includegraphics[width=3.5 in]{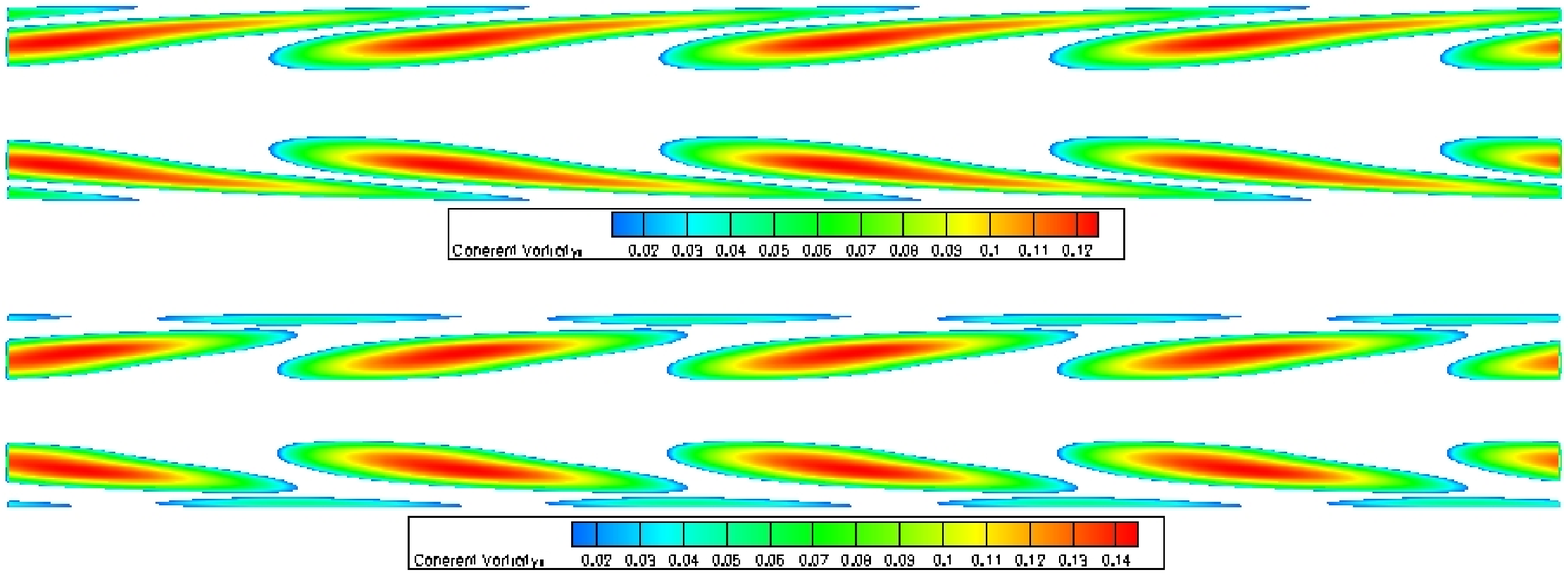}
\caption{Cross-section along the $r$-$z$ plane of coherent vorticity of (2,0,1) ring mode.  Top: (a)
  non-oscillated.  Bottom: (b) oscillated.  The vortex core shifts
  $y^+=7$ away from the wall.}
\label{201}
\end{figure}

\begin{figure}
\begin{tabular}{cc}
\begin{minipage}{1.75 in}
\includegraphics[width=1.75 in]{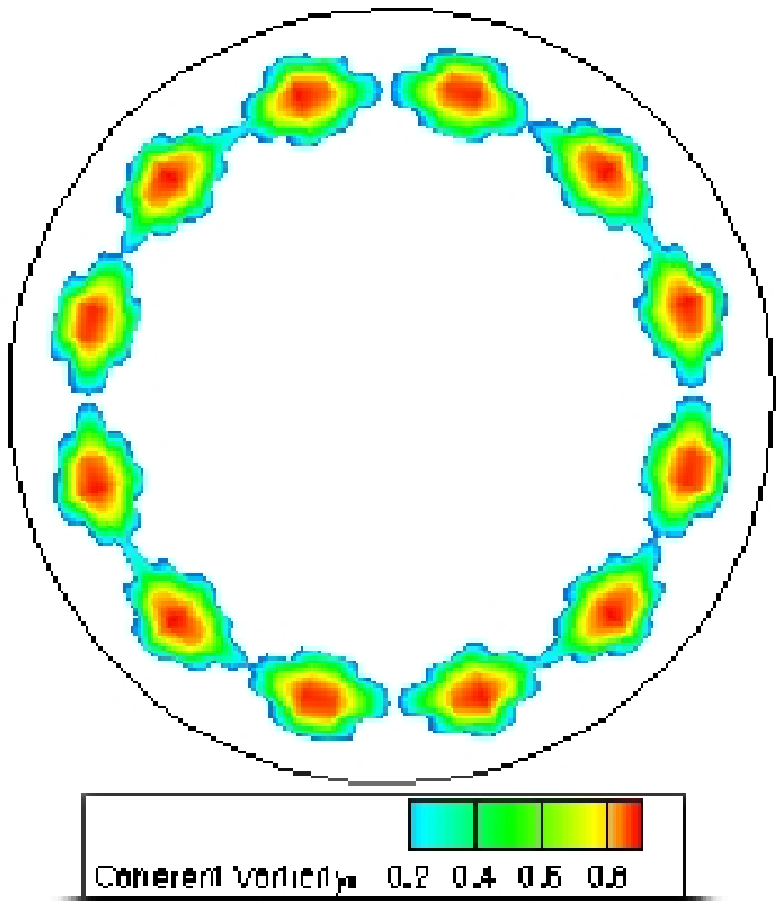}
\end{minipage}
&
\begin{minipage}{1.75 in}
\includegraphics[width=1.75 in]{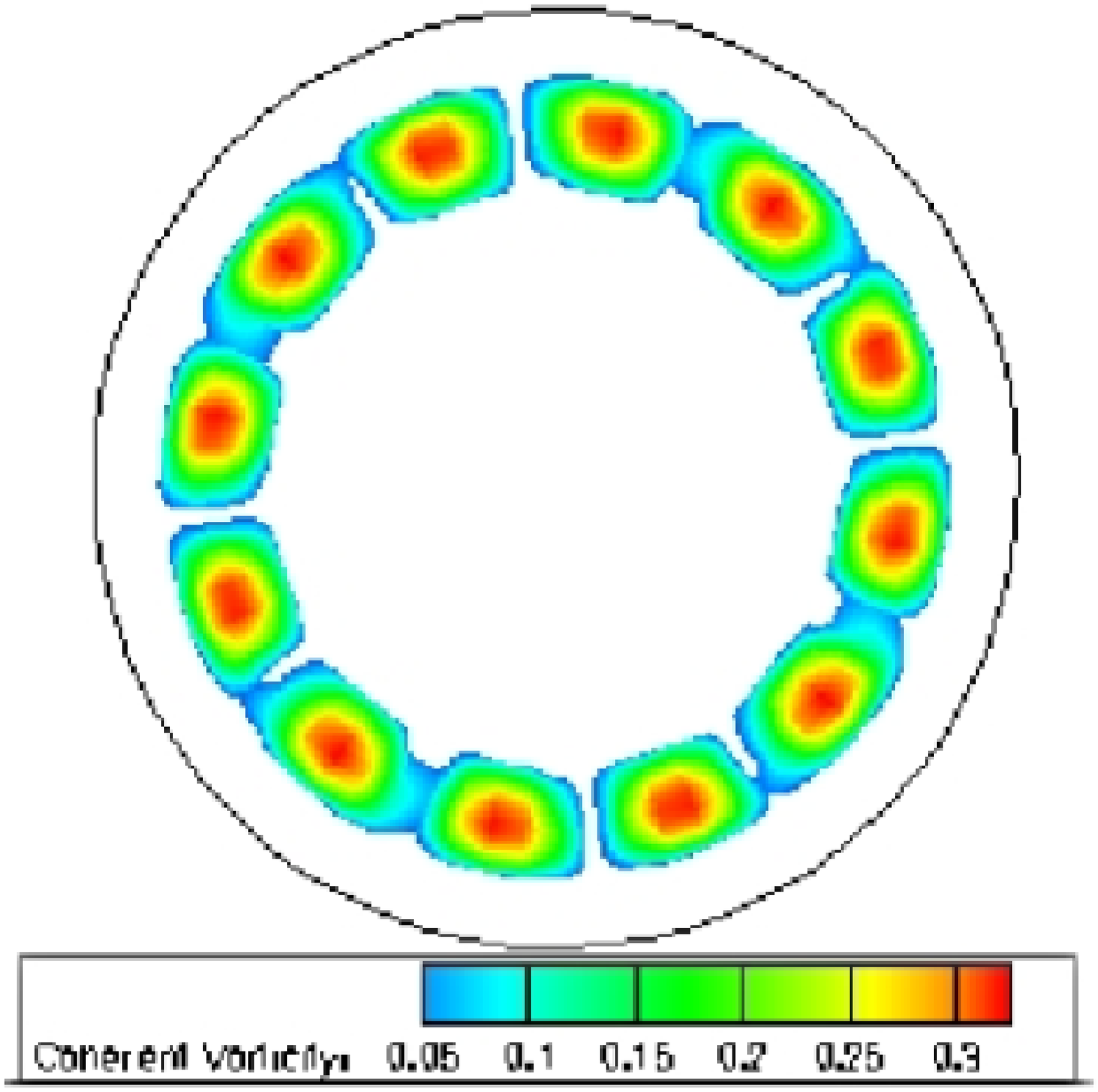}
\end{minipage}\\
\end{tabular}
\caption{Cross-section of coherent vorticity of (0,6,1) roll mode.  Left: (a)
  non-oscillated.  Right: (b) oscillated.  The vortex core shifts
  $y^+=10.1$ away from the wall.}
\label{061}
\end{figure}

\begin{figure}
\begin{tabular}{cc}
\begin{minipage}{1.75 in}
\includegraphics[width=1.75 in]{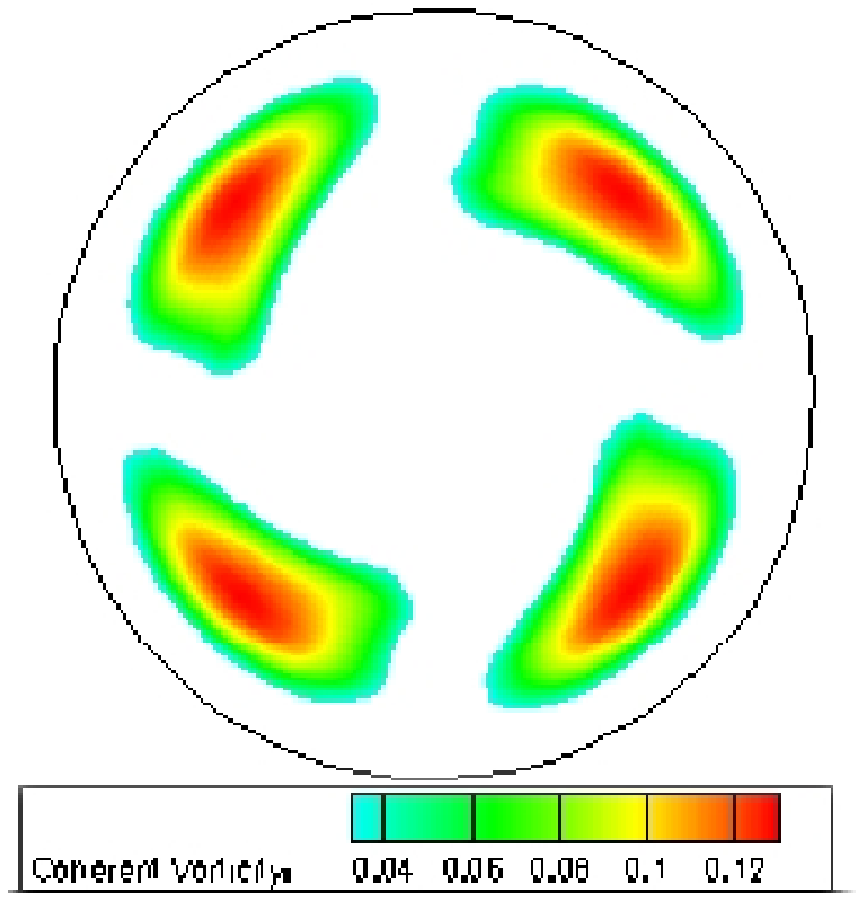}
\end{minipage}
&
\begin{minipage}{1.75 in}
\includegraphics[width=1.75 in]{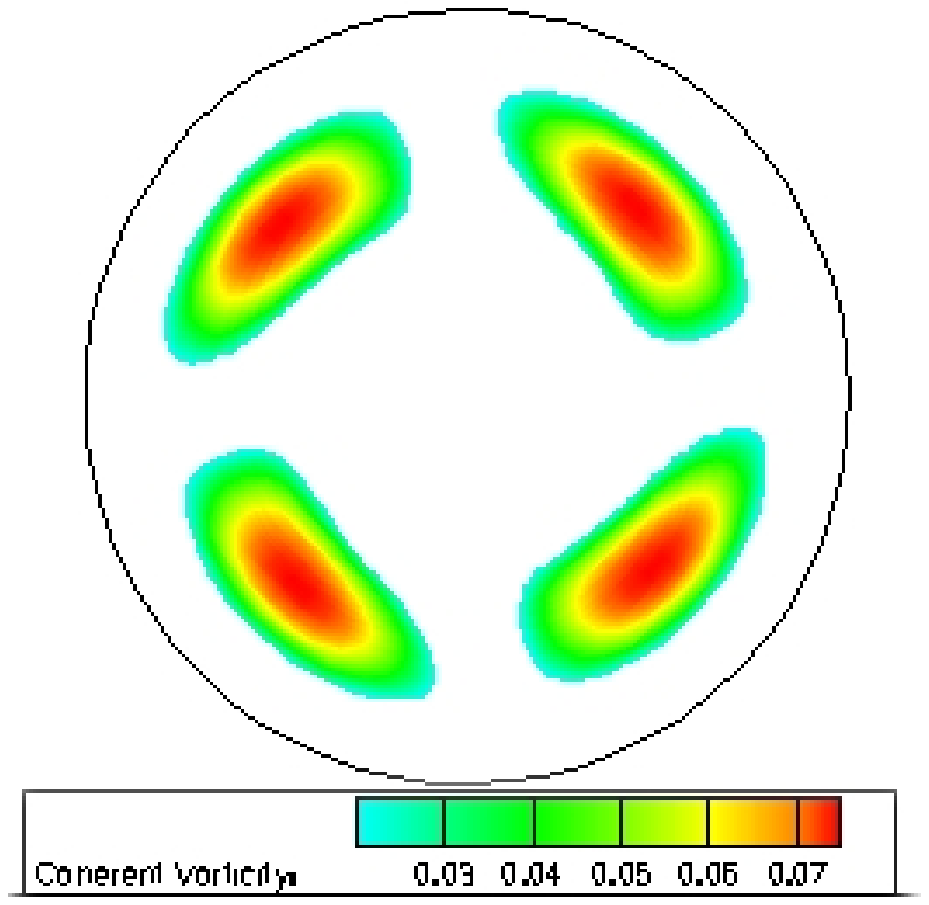}
\end{minipage}\\
\end{tabular}
\caption{Cross-section of coherent vorticity of (0,2,1) roll mode.  Left: (a)
  non-oscillated.  Right: (b) oscillated.  The vortex core shifts
  $y^+=7.8$ away from the wall.}
\label{021}
\end{figure}

The shifting of the structures away from the wall is shown for the most
energetic modes for each propagating subclass in Figures \ref{121} through \ref{021}.  These
structures, which represent the coherent vorticity of the four subclasses
of the propagating modes, are pushed towards the center of the
pipe, where the mean flow velocity is faster.  The location of the
coherent vortex core for these eight
modes are listed in Table \ref{corelocation}, showing a shift away
from the wall.  The only mode 
found not to follow this trend is the (1,0,1) mode, which
undergoes a major restructuring resulting in
its vortex core moving towards the wall.  

\begin{table}
\caption{Comparison of the measured location of the
  coherent vortex core for two modes from each propagating subclass in wall units ($y^+$) away from
  the pipe wall .
  Each class shifts away from the wall, consistent with the shift in
  velocity rms and Reynolds stress.  The (1,0,1) mode changes structure
  significantly in the oscillated case as seen in Figure \ref{101}, explaining the shift of its
  coherent vortex core towards the wall.}
\begin{tabular}{cccc}
\hline \hline
& \multicolumn{3}{c}{Coherent vortex core location (wall units)}
 \\
Mode &  Non-oscillated & Oscillated & shift\\
\hline
(1,2,1) & 41.8 & 48.6 & 6.8\\
(1,5,1) & 28.7 & 38.6 & 9.9\\
(2,2,1) & 45.6 & 56.2 & 11.0\\
(3,2,1) & 65.9 & 77.1 & 11.2 \\
(1,1,1) & 47.2 & 50.4 & 3.2\\
(2,1,1) & 53.9 & 99.8 & 45.9\\
(1,0,1) & 31.9 & 18.9 & -13.0\\
(2,0,1) & 58.4 & 65.8 & 7.4 \\
(0,6,1) & 28.9 & 39.0 & 10.1 \\
(0,2,1) & 45.1 & 52.9 & 7.8 \\
\hline \hline
\end{tabular}
\label{corelocation}
\end{table}

This faster advection explains the experimental results found by K.-S. Choi
\cite{choi_kwing} that showed a reduction in the duration and strength of
sweep events in a spanwise wall oscillated boundary layer of 78\% and
64\% respectively.  For this experiment, the flowrate was kept
constant, so the energy was reduced, whereas in our
case the mean pressure gradient was kept constant yielding virtually no
change in the energy of the propagating structures.  Also corroborating these results is the work by Prabhu et
al. \cite{prabhu} that examined the KL decomposition of controlled suction and
blowing to reduce drag in a channel.  They, too, found that the structures were
pushed away from the wall and that they had higher phase velocities.
Thirdly, the study by Zhou \cite{zhou_dongmei} is also consistent with
these results, as she found that any oscillation
in the streamwise direction reduces the effectiveness of the drag
reduction.  Any streamwise oscillation, even though it would still
push the structures away from the wall, would adversely effect the mean flow rate
profile resulting in an advection speed of the propagating waves that
is less than in the purely spanwise oscillated case.  

Thus, faster advection can be interpreted in
two fashions.  The first is in terms of the traveling wave.  The
shifting of the structures away from the wall into higher velocity mean
flow causes these structures to travel faster, giving them less
interaction time with the roll modes.  Less interaction time with the
roll modes means less energy transfer (less bursting), and due to
their fast decaying nature, their lifetime is reduced.  This reduced
lifetime means they have less time to generate Reynolds
stress, and therefore drag is reduced.  The second interpretation is in terms of the classically
observed hairpin and horseshoe vortices. \cite{theodorsen, robinson1991}  The pushing
of the KL structures away
from the wall is
equivalent to the vortices lifting and stretching away from the wall
faster.  This faster lifting and stretching process means that their
lifetime is shortened, again resulting is less time to generate Reynolds
stress, and therefore drag reduction occurs.

\section{CONCLUSIONS}
This work has shown, through a Karhunen-Lo\`{e}ve analysis, four major
consequences of spanwise wall oscillation on the turbulent pipe flow
structures.  
They are:  a shifting of rms velocities and Reynolds stress away
from the wall; a reduction in the dimension of the chaotic attractor
describing the turbulence; a decrease in the energy in the propagating modes as a whole with an increase in modes that transfer energy to the outside of the log layer; and a shifting of the propagating structures away from the
wall into higher speed flow resulting in faster advection and
shorter lifespans, providing less time to generate Reynolds stress and
therefore reducing drag.

The strength of the KL method is that
it yields global detail and structure without conditional sampling.  The ensemble was created out of evenly spaced
flowfields in time, as opposed to conditional sampling of the
flowfield with event detection such as bursts or sweeps, and the entire flowfield and time
history was studied.  Therefore, we argue that the overall mechanism of drag reduction through spanwise  
wall oscillation has been found.  Although a result of drag reduction
is the
decorrelation of the low 
speed streaks and the streamwise vortices, as found by previous researchers, this is an incomplete
description of the dynamics.  It is the lifting of the
turbulent structures away from the wall by the Stokes flow induced by
the spanwise wall oscillation that cause the reduction in the time and
duration of Reynolds stress
generating events, resulting in drag reduction.  In addition, this
dynamical description encompasses other methods of drag reduction such as
suction and blowing, active control, and ribblets, \cite{prabhu} establishing it as a
contender for a universal theory of drag reduction.

\section*{ACKNOWLEDGMENTS}
This research was supported in part by the National Science Foundation
through TeraGrid resources provided by the San Diego Supercomputing
Center, and by Virginia Tech through their Terascale Computing
Facility, System X.   We gratefully acknowledge many useful interactions with Paul Fischer
and for the use of his
spectral element algorithm.

\bibliography{references}

\end{document}